\documentclass[12pt]{article}
\usepackage{amsmath,amssymb,amsthm,amsfonts,bm,mathrsfs,float,algorithm,algorithmic,caption,subcaption,setspace}
\usepackage{avant,array,geometry,fontenc,inputenc,natbib,hyperref,multirow,enumerate,enumitem,rotating,booktabs,color,latexsym,times,stmaryrd,xr}
\makeatletter
\@ifundefined{date}{}{\date{}}

\newcolumntype{L}[1]{>{\raggedright\let\newline\\\arraybackslash\hspace{0pt}}m{#1}}
\newcolumntype{C}[1]{>{\centering\let\newline\\\arraybackslash\hspace{0pt}}m{#1}}
\newcolumntype{R}[1]{>{\raggedleft\let\newline\\\arraybackslash\hspace{0pt}}m{#1}}

\floatstyle{ruled}
\newfloat{algorithm}{tbp}{loa}
\providecommand{\algorithmname}{Algorithm}
\floatname{algorithm}{\protect\algorithmname}
\@ifundefined{definecolor}{color}{}

\newtheorem{pn}{{\bf Proposition}}

\newtheorem{thm}{{\bf Theorem}}

\newtheorem{definition}{{\bf Definition}}

\global\long\def\mbA{\boldsymbol{A}}%
\global\long\def\mbD{\boldsymbol{D}}%
\global\long\def\mbf{\boldsymbol{f}}%
\global\long\def\mbg{\boldsymbol{g}}%
\global\long\def\mbG{\boldsymbol{G}}%
\global\long\def\mbI{\boldsymbol{I}}%
\global\long\def\mbO{\boldsymbol{O}}%
\global\long\def\mbP{\boldsymbol{P}}%
\global\long\def\mbv{\boldsymbol{v}}%
\global\long\def\mbX{\boldsymbol{X}}%
\global\long\def\mbY{\boldsymbol{Y}}%
\global\long\def\mbz{\boldsymbol{z}}%
\global\long\def\mbZ{\boldsymbol{Z}}%
\global\long\def\bolgamma{\boldsymbol{\gamma}}%
\global\long\def\bolGamma{\boldsymbol{\Gamma}}%
\global\long\def\bolDelta{\boldsymbol{\Delta}}%
\global\long\def\bolSigma{\boldsymbol{\Sigma}}%
\global\long\def\bolPhi{\boldsymbol{\Phi}}%
\global\long\def\hatbolGamma{\widehat{\boldsymbol{\Gamma}}}%
\global\long\def\hatbolDelta{\widehat{\boldsymbol{\Delta}}}%
\global\long\def\mbbR{\mathbb{R}}%
\global\long\def\calS{\mathcal{S}}%
\global\long\def\mbzero{\boldsymbol{0}}%
\global\long\def\tilbolPhi{\widetilde{\bolPhi}}%
\global\long\def\E{\mathrm{E}}%
\global\long\def\Cov{\mathrm{cov}}%
\global\long\def\Corr{\mathrm{corr}}%
\global\long\def\spn{\mathrm{span}}%
 \global\long\def\tilbolDelta{\boldsymbol{\widetilde{\Delta}}}%

\newcommand{\independent}{\protect\mathpalette{\protect\independenT}{\perp}}
\def\independenT#1#2{\mathrel{\rlap{$#1#2$}\mkern2mu{#1#2}}}

\global\long\def\LA{\mathrm{LA}}%
\global\long\def\GLA{\mathrm{GLA}}%

\begin{document}

\title{\Large{\textbf{Generalized Liquid Association Analysis \\ 
for Multimodal Data Integration}}}
\author{
\bigskip
Lexin Li$^\dag$, Jing Zeng$^\ddag$, and Xin Zhang$^\ddag$ \thanks{The authors contributed equally to this work and are listed in alphabetical order.} \\
\normalsize{\textit{$^\dag$University of California at Berkeley and $^\ddag$Florida State University}}
}
\date{}
\maketitle

\begin{abstract}
Multimodal data are now prevailing in scientific research. A central question in multimodal integrative analysis is to understand how two data modalities associate and interact with each other given another modality or demographic variables. The problem can be formulated as studying the associations among three sets of random variables, a question that has received relatively less attention in the literature. In this article, we propose a novel generalized liquid association analysis method, which offers a new and unique angle to this important class of problem of studying three-way associations. We extend the notion of liquid association of \citet{li2002LA} from the univariate setting to the sparse, multivariate, and  high-dimensional setting. We establish a population dimension reduction model, transform the problem to sparse Tucker decomposition of a three-way tensor, and develop a higher-order orthogonal iteration algorithm for parameter estimation. We derive the non-asymptotic error bound and asymptotic consistency of the proposed estimator, while allowing the variable dimensions to be larger than and diverge with the sample size. We demonstrate the efficacy of the method through both simulations and a multimodal neuroimaging application for Alzheimer's disease research. 
\end{abstract}


\newpage
\baselineskip=20pt

%%%%%%%%%%%%%%%%%%%%%%%%%%%%%%%%%%%%%%%%%%%%%%%%%%%
\section{Introduction}
\label{sec:introduction}

%%%%%%%%%%%%%%%%%%%%%%%%%%%%%%%%%%%%%%%%%%%%%%%%%%%
\subsection{Motivation and problem formulation}
\label{subsec:motivation}

Multimodal data are now prevailing in scientific research, where different types of data are acquired for a common set of experimental subjects. One example is multi-omics, where different genetic information such as gene expressions, copy number alternations, and methylation changes are jointly collected for the same biological samples \citep{richardson2016statistical}. Another example is multimodal neuroimaging, where distinct brain characteristics including brain structure, function, and chemical constituents are simultaneously measured for the same study subjects \citep{Uludag2014}. Integrative analysis of multimodal data aggregates such diverse and often complementary information, and consolidates knowledge across multiple data modalities. 

In this article, we aim to address a question of central interest in multimodal integrative analysis, i.e., to understand how different data modalities associate and interact with each other given other modalities or covariates. This problem is of a broad scientific interest; for instance, it is useful to understand how gene expressions and microRNA levels interact given the severity of ovarian cancer and other demographical variables \citep{Cai2016}, or how gene expressions and comparative genomic hybridization measures interact given the progression of breast cancer and demographics \citep{mai2019iterative}. Our motivation is a multimodal positron emission tomography (PET) study for Alzheimer's disease (AD). Amyloid-beta and tau are two hallmark proteins of AD, both of which can be measured in vivo by PET imaging using different nuclear tracers. The two proteins are closely associated in terms of spatial patterns of their accumulations, and such association patterns are believed to be affected by the subject's age \citep{Braak1991}. Nevertheless, their specific age-dependent regional associations remain unclear. The data we study involve 81 elderly subjects, each receiving two PET scans that measure the depositions of amyloid-beta and tau, respectively. Each PET modality is represented by a vector of protein measurements at a set of brain regions of interest, with 60 regions for amyloid-beta, and 26 for tau. Our goal is to find how and where in the brain the associations of the two proteins are the most contrastive as the subject's age varies. 

This problem can be formulated statistically as studying the associations of two sets of random variables $\mbX \in \mbbR^{p_{1}}$ and $\mbY \in \mbbR^{p_{2}}$ conditional on the third set of random variables $\mbZ \in \mbbR^{p_{3}}$. In our motivation example, $\mbX$ denotes the amyloid-beta PET imaging with $p_1=60$, $\mbY$ denotes the tau PET imaging with $p_2=26$, and $\mbZ$ denotes the subject's age with $p_3=1$. Meanwhile, in plenty of multimodal applications, $\mbX, \mbY, \mbZ$ can all be high-dimensional, and their dimensions can be even larger than the sample size. For instance, in imaging genetics \citep{Zhu2017review}, $\mbX, \mbY$ can represent different imaging modalities, whose dimensions can be in hundreds, and $\mbZ$ can denote the genetic information, whose dimension can be in tens of thousands or more. In high-dimensional data analysis, it is common to postulate that the data information can be sufficiently captured by some low-dimensional representations, and most often, some linear combinations of the originally high-dimensional variables \citep{Cook2007}. Adopting this view, our question can be formulated as seeking linear combinations of $\mbX$ and linear combinations of $\mbY$ whose conditional associations given $\mbZ$ are the most contrastive. In other words, we seek linear combinations of $\mbX$ and $\mbY$ that change the most as $\mbZ$ changes.

%%%%%%%%%%%%%%%%%%%%%%%%%%%%%%%%%%%%%%%%%%%%%%%%%%%
\subsection{Related work}

There has been a rich statistical literature studying the associations between two sets of multivariate variables $\mbX$ and $\mbY$. A well studied and commonly used family of methods are canonical correlation analysis (CCA) and its variants \citep[among others]{witten2009penalized, Gao2015, li2018general, shu2019dcca,  mai2019iterative}. CCA explores the symmetric relations between $\mbX$ and $\mbY$, and looks for pairs of linear combinations that are most correlated. This goal, however, is different from ours, as the highly correlated linear combinations of $\mbX$ and $\mbY$ are not necessarily the ones that are the most contrastive. For instance, a pair of linear combinations of $\mbX$ and $\mbY$ can be highly correlated, while this correlation remains a constant as the value of $\mbZ$ varies, and as such they are not the target of our problem. We later numerically compare our method with CCA to further demonstrate their differences. Another popular family of methods are sufficient dimension reduction (SDR), which looks for linear combinations of $\mbX$ that capture full regression information of $\mbY$ given $\mbX$; see the recent book of \citet{Li2018SDRBook} for a review of this topic. Later we show that our proposed method is connected to several SDR methods, including principal Hessian directions \citep{li1992principal, cook1998principal, tang2020high}, and partial and groupwise sufficient dimension reduction \citep{partialSDR, groupSDR}. However, the goals of the two are utterly different. Whereas SDR studies asymmetric relations of $\mbY$ conditioning on $\mbX$, we seek symmetric relations between $\mbX$ and $\mbY$ conditioning on the third set of variables $\mbZ$, in that the roles of $\mbX$ and $\mbY$ are interchangeable, but not with the role of $\mbZ$. 

Compared to the setting of two sets of variables, there have been much fewer statistical methods studying the associations among three sets of multivariate variables in the form of $\mbX$ and $\mbY$ given $\mbZ$. In his groundbreaking work, \citet{li2002LA} proposed a novel three-way interaction metric, termed \emph{liquid association}, that measures the extent to which the association of a pair of random variables depends on the value of a third variable. He showed that this metric is particularly useful in discovering co-expressed gene pairs that are regulated by another gene. However, \citet{li2002LA} only considered the scenario where all three variables $X, Y, Z$ are one-dimensional. \citet{li2004PLA} extended the notion of liquid association to the scenario of a multivariate $\mbX$ and a scalar $Z$, and sought two linear combinations $\bolgamma_1^T \mbX$ and  $\bolgamma_2^T \mbX$ such that $\Corr(\bolgamma_1^T\mbX,\bolgamma_2^T\mbX | Z)$ varies the most with $Z$. \citet{ho2011MLA} and \citet{Yu2018} developed some modified versions of liquid association, but still focused on the one-dimensional $X, Y, Z$ scenario. Relatedly, \citet{chen2011penalized} proposed a bivariate conditional normal model to identify the variables that regulate the co-expression patterns between two genes. That corresponds to the scenario with a scalar $X$, a scalar $Y$ and a multivariate $\mbZ$. \citet{Abid2018} proposed contrastive principal component analysis for a multivariate $\mbX$ and a binary scalar $Z$, which sought linear combinations of $\mbX$ that have the largest changes in the conditional variance given $Z=0$ versus $Z=1$. Moreover, \citet{Lock2013, li2017incorporating} developed a class of matrix and tensor factorization methods, which aimed to decompose the multimodal data into the components that capture joint variation shared across modalities, and the components that characterize modality-specific variation. Their goal is again different from ours, as their methods did not target the conditional distribution of $\mbX, \mbY$ given $\mbZ$. Finally, \citet{XiaLi2019} analyzed a similar dataset as our motivation example, but tackled a totally different problem. They studied hypothesis testing of covariance between the two multivariate PET measurements, and worked on the residuals after regressing out the age effect, which involves no conditioning of any third set of variables.

%%%%%%%%%%%%%%%%%%%%%%%%%%%%%%%%%%%%%%%%%%%%%%%%%%%
\subsection{Proposal and contributions}

In this article, we study the three-way association among multivariate $\mbX, \mbY, \mbZ$, and seek a set of linear combinations of $\mbX$ and $\mbY$ that has a varying association as $\mbZ$ varies. We generalize the notion of liquid association of \citet{li2002LA} from the univariate case to the multivariate case, and develop a population dimension reduction framework for three-way association analysis. Our extension is far from trivial though, resulting in a completely different estimation method and the associated asymptotic theory. For the estimation, we transform the liquid association analysis to the problem of sparse Tucker decomposition of a three-way tensor, and introduce sparsity for the linear combinations to improve the interpretability. We then develop a higher-order orthogonal iteration algorithm for parameter estimation. For the theory, we establish a population model that is essential for the study of statistical properties. We then derive the error bound and consistency, while allowing the variable dimensions $p_1, p_2, p_3$ to be larger than and to diverge to infinity along with the sample size $n$. As a result, our proposal makes some useful contributions from both the scientific and statistical perspectives. 

Scientifically, characterizing the associations between different modalities given other modalities or covariates is of crucial importance for multimodal integrative analysis. However, there is almost no existing statistical solution available for this type of problem, especially when all the modalities involved are high-dimensional. Our proposal offers a unique angle for this important class of problem. As an illustration, for our multimodal PET study, understanding the dynamic patterns between amyloid-beta and tau with respect to age would offer insight about how pathological proteins of Alzheimer's disease interact in the aging human brain.     

Statistically, our proposal of generalized liquid association analysis makes a useful addition to the toolbox of association analysis of more than two sets of variables. Moreover, our method involves sparse tensor decomposition, which is itself of independent interest. Tensor data analysis is gaining increasing attention in recent years \citep[among others]{kolda2009tensor, ZhouLiZhu2013, sun2017provable, bi2018tensor, tang2019tensor, zhang2019optimal, Hao2020}; see also \citet{BiQu2020} for a review of tensor analysis in statistics. Nevertheless, our proposal differs in several ways. In particular, our sparse tensor decomposition algorithm is related to some recent singular value decomposition (SVD) type solutions for matrix denoising \citep{yang2016rate} and tensor denoising \citep{zhang2019optimal}, in that they share a similar iterative hard thresholding SVD scheme. However, our algorithm is tailored to the tensor parameter estimation with more flexible initialization and tuning. As a result, our theoretical analysis differs considerably from the denoising problems. Whereas both \citet{yang2016rate} and \citet{zhang2019optimal} achieved the minimax optimal estimation for their denoising problems, we establish the dimension reduction subspace recovery consistency, variable selection consistency, and tensor parameter estimation consistency. Our rate of convergence matches the optimal rate in previous works, and all the consistency results are established in the ultrahigh dimensional setting of $s\log(p)=o(n)$, where $s=s_1s_2s_3$ and $p=p_1p_2p_3$ are the products of the number of nonzero entries and dimensions, respectively, in $\mbX$, $\mbY$ and $\mbZ$.  Our  theoretical development is highly non-trivial, and may be of independent interest for future research involving tensor parameter estimation in a statistical model with i.i.d.\ data. In a sense, our work further broadens the scope of higher-order sparse SVD and tensor analysis.

The rest of the article is organized as follows. Section \ref{sec:model} develops the concept of generalized liquid association, and the corresponding population model of generalized liquid association analysis. Section \ref{sec:estimation} introduces the estimation algorithm, and Section \ref{sec:theory} establishes the theoretical guarantees. Section \ref{sec:sim} presents the simulations, and Section \ref{sec:realdata} revisits the multimodal PET study. Section \ref{sec:discussion} concludes the paper with a discussion, and the Supplementary Appendix collects all technical proofs and additional numerical results.

%%%%%%%%%%%%%%%%%%%%%%%%%%%%%%%%%%%%%%%%%%%%%%%%%%%
\section{Generalized Liquid Association Analysis}
\label{sec:model}

We first generalize the concept of liquid association from the univariate case to the multivariate case. The conceptual generalization itself is straightforward. Nevertheless, it motivates us to develop a dimension reduction model, along with an optimization formulation, that connects to the problem of tensor decomposition. We show that the solution to this  optimization problem is indeed the low-dimensional representation in the generalized liquid association that we seek, and this result holds without having to require the normality assumption. Our method provides essentially a new dimension reduction framework for three-way association analysis.

%%%%%%%%%%%%%%%%%%%%%%%%%%%%%%%%%%%%%%%%%%%%%%%%%%%
\subsection{Generalized liquid association}

We begin with a brief review of the concept of liquid association (LA) proposed by \citet{li2002LA} for the univariate case. We then extend this notion to the multivariate case. 

Suppose $X, Y, Z$ are random variables with mean zero and variance one. Define $g(z) = \E(XY | Z = z) : \mbbR \mapsto \mbbR$. \citet{li2002LA} defined the liquid association of $X$ and $Y$ given $Z$ as, 
\begin{equation*}
\LA(X,Y\mid Z) = \E\left\{ \dfrac{d g(Z)}{d Z} \right\} = \E\left\{ \dfrac{d}{dZ}\E(XY\mid Z) \right\},
\end{equation*}
When $Z$ follows a standard normal distribution, by Stein's Lemma \citep{stein1981}, we have, 
\begin{equation*}
\LA(X,Y\mid Z)=\E\left\{ g(Z)Z \right\} = \E(XYZ).
\end{equation*}
Intuitively, $\LA(X,Y\mid Z)$ characterizes the change of the association of $X$ and $Y$ conditioning on $Z$ through $g(z)$, and the normality condition connects this quantity with the simple unconditional expectation $\E(XYZ)$. In practice, the univariate $Z$ is transformed to standard normal using the normal score transformation, and the LA measure is estimated by the sample mean $\E(XYZ)$. \citet{li2004PLA} considered an extension of LA to a multivariate $\mbX \in \mbbR^{p_1}$ and a scalar $Z$, by looking for two linear combinations, such that $\LA(\bolgamma_1^T \mbX, \bolgamma_2^T \mbX | Z) = \bolgamma_1^T \E(\mbX\mbX^{T}Z) \bolgamma_2$ is maximized. It has a close-form solution that $\bolgamma_1 = (\mbv_{1}+\mbv_{p})/\sqrt{2}$, $\bolgamma_2 = (\mbv_{1}-\mbv_{p})/\sqrt{2}$, where $\mbv_1$ and $\mbv_p$ are the eigenvectors of the matrix $\E(\mbX\mbX^{T}Z)\in\mbbR^{p\times p}$ with the largest and smallest eigenvalues. 

We next extend the concept of liquid association to the multivariate case, where $\mbX \in \mbbR^{p_{1}}$, $\mbY \in \mbbR^{p_{2}}$, and $\mbZ\in\mbbR^{p_{3}}$. Without loss of generality, suppose each variable entry in $\mbX$, $\mbY$, and $\mbZ$ are standardized with mean zero and variance one. Define 
\begin{equation*}
\mbg(\mbz) = \E(\mbX\mbY^{T} \mid \mbZ = \mbz) : \mbbR^{p_{3}} \mapsto \mbbR^{p_{1} \times p_{2}}.
\end{equation*}
We introduce the generalized liquid association measure, which is a three-way tensor.

\begin{definition}\label{def:LA} 
The generalized liquid association (GLA) of $\mbX$ and $\mbY$ with respect to $\mbZ$ is,
\begin{equation*}
\bolPhi = \GLA(\mbX,\mbY \mid \mbZ) = \E\left\{ \dfrac{d}{d\mbZ}\mbg(\mbZ) \right\} \in \mbbR^{p_{1} \times p_{2} \times p_{3}}.
\end{equation*}
\end{definition}

When $\mbZ$ follows a multivariate normal distribution, by the multivariate version of Stein's Lemma \citep[Lemma 1]{Liu1994}, we have the following property regarding $\bolPhi$. 

\begin{pn} \label{pn:multivariateLA} 
If $\mbZ \sim \mathrm{Normal}(0, \bolSigma_{\mbZ})$, then $\bolPhi = \E(\mbX \circ \mbY \circ \mbZ) \times_{3} \bolSigma_{\mbZ}^{-1}$, where $\circ$ denotes the outer product, and $\times_{3}$ denotes the mode-3 product between a tensor and a matrix.
\end{pn}

\noindent
This conceptual extension from univariate to multivariate variables is straightforward. However, we recognize that all $\mbX, \mbY$ and $\mbZ$ can be high-dimensional such that $p_1, p_2, p_3 > n$, and the dimension of $\bolPhi$ is the product $p_1 p_2 p_3$, which can be ultrahigh-dimensional. Besides, it involves the inversion of a potentially high-dimensional covariance matrix $\bolSigma_{\mbZ}$, which makes a direct calculation or any operation on $\bolPhi$ difficult, if not completely infeasible. Finally, the normality assumption can be restrictive, and there may be no simple way to transform a multivariate or high-dimensional $\mbZ$ to follow an approximate normal distribution. Next, we develop a dimension reduction model for $\bolPhi$, which reduces the dimensionality, avoids $\bolSigma_{\mbZ}^{-1}$, and improves the interpretability of the result. We also examine the normality assumption carefully, and show that it is \emph{not} absolutely necessary for our generalized liquid association analysis.

%%%%%%%%%%%%%%%%%%%%%%%%%%%%%%%%%%%%%%%%%%%%%%%%%%%
\subsection{Dimension reduction model for three-way association}

We next propose a dimension reduction model for three-way association. Our goal is to seek the linear combinations of $\mbX$ and $\mbY$ that change the most as $\mbZ$ or its linear combinations change. 

Specifically, we first postulate that the matrix $\mbg(\mbz)=\E(\mbX\mbY^{T} | \mbZ=\mbz) \in \mbbR^{p_{1}\times p_{2}}$ varies within a low-dimensional subspace for all values of $\mbz$, in that 
\begin{equation*}
\mbg(\mbz) = \E(\mbX\mbY^{T} | \mbZ = \mbz) = \bolGamma_{1} \mbf_z(\mbz) \bolGamma_{2}^{T}, 
\end{equation*}
for some semi-orthogonal basis matrices $\bolGamma_{k}\in\mbbR^{p_{k}\times r_{k}}, k=1,2$, and some latent function $\mbf_z:\mbbR^{p_{3}}\mapsto\mbbR^{r_{1}\times r_{2}}$. This implies that the linear combinations $\bolGamma_1^{T} \mbX$ and $\bolGamma_2^{T} \mbY$ capture all the variations in the first two modes of the generalized liquid association tensor $\bolPhi$. Next, we further assume that $\mbg(\mbz)$ depends on $\mbZ$ only through a few linear combinations $\bolGamma_{3}^{T}\mbZ$ of $\mbZ$, for some semi-orthogonal basis matrix $\bolGamma_{3}\in\mbbR^{p_{3}\times r_{3}}$. Putting these two dimension reduction structures together, we obtain our dimension reduction model for the general three-way association analysis: 
\begin{equation} \label{XYZsub}
\mbg(\mbz) = \E(\mbX\mbY^{T} \mid \mbZ = \mbz) = \bolGamma_{1} \mbf(\bolGamma_{3}^{T}\mbz) \bolGamma_{2}^{T}, 
\end{equation}
for some latent function $\mbf:\mbbR^{r_{3}}\mapsto\mbbR^{r_{1}\times r_{2}}$. This model is to serve as the basis for our subsequent generalized liquid association analysis. Later, we further introduce sparsity to $(\bolGamma_{1}, \bolGamma_{2}, \bolGamma_{3})$ to improve the interpretability of the linear combinations we identify in model \eqref{XYZsub}.

%%%%%%%%%%%%%%%%%%%%%%%%%%%%%%%%%%%%%%%%%%%%%%%%%%%
\subsection{Generalized liquid association analysis via tensor decomposition}

We propose to estimate the linear combination coefficient $\bolGamma_{k}$ in the dimension reduction model \eqref{XYZsub}, or more accurately, the subspace $\spn(\bolGamma_k)$ spanned by the columns of $\bolGamma_{k}$, $k=1,2,3$, by solving the following optimization problem, 
\begin{equation} \label{tuckerObj2}
\mathrm{minimize}_{\mbG_{1},\mbG_{2},\mbG_{3}} \left\Vert \bolDelta - \bolDelta \times_{1}\mbP_{\mbG_{1}}\times_{2}\mbP_{\mbG_{2}}\times_{3}\mbP_{\mbG_{3}} \right\Vert_{F}^{2}, 
\end{equation}
where $\bolDelta = \E(\mbX \circ \mbY \circ \mbZ) \in \mbbR^{p_{1} \times p_{2} \times p_{3}}$, $\mbG_{k}\in\mbbR^{p_{k}\times r_{k}}$ is a semi-orthogonal matrix, $\mbP_{\mbG_k} = \mbG_k(\mbG_k^{T}\mbG_k)^{-1}\mbG_k^{T}$ is the projection onto the subspace $\spn(\mbG_k)$, and $\times_{k}$ is the mode-$k$ product between a tensor and a matrix, $k=1,2,3$. We first note that the optimization in \eqref{tuckerObj2} is actually the well-known tensor Tucker decomposition \citep{kolda2009tensor}. Let $(\widehat{\mbG}_{1}, \widehat{\mbG}_{2}, \widehat{\mbG}_{3})$ denote the minimizer of \eqref{tuckerObj2}. We next carefully study the connections between $(\widehat{\mbG}_{1}, \widehat{\mbG}_{2}, \widehat{\mbG}_{3})$, the linear combination coefficient $(\bolGamma_{1}, \bolGamma_{2}, \bolGamma_{3})$ in model \eqref{XYZsub}, and the GLA measure $\bolPhi$ in Definition \ref{def:LA}.

Toward that end, we introduce an intermediate optimization problem, 
\begin{equation} \label{tuckerObj}
\mathrm{minimize}_{\mbG_{1},\mbG_{2},\mbG_{3}} \left\Vert \bolPhi - \bolPhi \times_{1}\mbP_{\mbG_{1}}\times_{2}\mbP_{\mbG_{2}}\times_{3}\mbP_{\mbG_{3}} \right\Vert_{F}^{2}.
\end{equation}
Let $(\widetilde{\mbG}_{1}, \widetilde{\mbG}_{2}, \widetilde{\mbG}_{3})$ denote the minimizer of \eqref{tuckerObj}. Then $(\widetilde{\mbG}_{1}, \widetilde{\mbG}_{2}, \widetilde{\mbG}_{3})$ is also the solution to the maximization problem, 
\begin{equation*}
\mathrm{maximize}_{\mbG_{1},\mbG_{2},\mbG_{3}} \left\Vert \bolPhi \times_{1}\mbP_{\mbG_{1}}\times_{2}\mbP_{\mbG_{2}}\times_{3}\mbP_{\mbG_{3}} \right\Vert_{F}^{2}. 
\end{equation*}
In other words, solving \eqref{tuckerObj} helps find the linear combinations of $\mbX$ and $\mbY$ whose generalized liquid association given some linear combination of $\mbZ$ is maximized. In this sense, it achieves our goal of finding the most contrastive associations of $\mbX$ and $\mbY$ given $\mbZ$. 

The next theorem characterizes the relations among $(\widetilde{\mbG}_{1}, \widetilde{\mbG}_{2}, \widetilde{\mbG}_{3})$, $(\widehat{\mbG}_{1}, \widehat{\mbG}_{2}, \widehat{\mbG}_{3})$, and $(\bolGamma_{1}, \bolGamma_{2},$ $\bolGamma_{3})$. Basically, it says that minimizing \eqref{tuckerObj2} and minimizing \eqref{tuckerObj} give the same estimates of $\bolGamma_1$ and $\bolGamma_2$ in model \eqref{XYZsub}, in the sense that they span the same subspaces. Furthermore, if $\bolGamma_{3}^{T}\mbZ$ is normally distributed, then the estimates of $\bolGamma_3$ under the two minimization problems differ by a rotation. 

\begin{thm} \label{thm: optim} 
Suppose model \eqref{XYZsub} holds. Then,
\begin{enumerate}[label={(\alph*)}]
\item $\spn(\bolGamma_{k}) = \spn(\widetilde{\mbG}_{k}) = \spn(\widehat{\mbG}_{k})$, for $k=1,2$;
\item $\spn(\bolGamma_{3}) = \spn(\widetilde{\mbG}_{3}) = \bolSigma_{\mbZ}^{-1}\spn(\widehat{\mbG}_{3})$, if $\bolGamma_{3}^{T}\mbZ$ is normally distributed. 
\end{enumerate}
\end{thm}

Theorem \ref{thm: optim} justifies that we can achieve our goal of finding the linear combinations of $\mbX$ and $\mbY$ that are the most contrastive given $\mbZ$ through the optimization problem \eqref{tuckerObj2}, with two crucial implications. First, \eqref{tuckerObj2} only involves the three-way tensor $\bolDelta$, but does not require the inversion of the  potentially high-dimensional matrix $\bolSigma_{\mbZ}^{-1}$ as in $\bolPhi$ in \eqref{tuckerObj}. Second, and perhaps more importantly, we do not have to require the normality of neither $\mbZ$ nor $\bolGamma_{3}^{T}\mbZ$. This is because, regardless of the distribution of $\bolGamma_{3}^{T}\mbZ$, the minimizer $(\widehat{\mbG}_{1}, \widehat{\mbG}_{2})$ from \eqref{tuckerObj2} is the same as the minimizer $(\widetilde{\mbG}_{1}, \widetilde{\mbG}_{2})$ from \eqref{tuckerObj}, and thus they share the same interpretation. Only if we aim to recover $\bolGamma_3$, then we need both $\bolSigma_{\mbZ}^{-1}$ and the normality of $\bolGamma_{3}^{T}\mbZ$. However, we argue that, in our generalized liquid association analysis, our primary goal is to find the linear combinations of $\mbX$ and $\mbY$ that change the most given $\mbZ$. As such, we are more interested in the estimation of $\bolGamma_1$ and $\bolGamma_2$, whereas the estimation of $\bolGamma_3$ provides additional dimension reduction, but is, relatively speaking, of less interest. Our proposed dimension reduction model \eqref{XYZsub} essentially serves as a bridge that connects the two optimization problems \eqref{tuckerObj2} and \eqref{tuckerObj}, which in turn connects the Tucker decomposition formulation in \eqref{tuckerObj2} with the generalized liquid association measure $\bolPhi$. 

Finally, we remark that, our proposal is similar in spirit to an SDR method, i.e., the principal Hessian directions \citep{li1992principal}. It was also derived based on Stein's Lemma, but was shown to be useful for finding low-dimensional representations in graphics \citep{cook1998principal}, and for detecting interaction terms in regressions \citep{tang2020high}, even without the normality assumption.

%%%%%%%%%%%%%%%%%%%%%%%%%%%%%%%%%%%%%%%%%%%%%%%%%%%
\section{Sparse Tensor Estimation}
\label{sec:estimation}

Tucker decomposition is usually solved by a tensor SVD type algorithm, for example, a higher-order orthogonal iteration (HOOI) algorithm, which was first proposed by \citet{HOOI} and later studied in statistical models \citep[e.g.,][]{zhang2018tensor,luo2020sharp}. Next, we develop an iterative algorithm to solve \eqref{tuckerObj2}. We further introduce sparsity in this decomposition to improve the interpretability of the result. 

For $n$ i.i.d.\ data observations $\{ \mbX_i, \mbY_i, \mbZ_i, i=1,\ldots,n\}$, without loss of generality, we assume the data is centered, so that $\sum_{i=1}^{n}\mbZ_{i}=0$. The centering of $\mbX$ and $\mbY$ is not required, but for simplicity, we assume $\sum_{i=1}^{n} \mbX_{i} = \sum_{i=1}^{n}\mbY_{i} = 0$ as well. Then the sample estimator of $\bolDelta$ is simply $\tilbolDelta = n^{-1} \sum_{i=1}^{n}\mbX_{i}\circ\mbY_{i}\circ\mbZ_{i} \in \mbbR^{p_{1} \times p_{2} \times p_{3}}$. Following the dimension reduction model \eqref{XYZsub} and the optimization problem \eqref{tuckerObj2}, $\tilbolDelta$ admits a Tucker tensor decomposition structure, which can be solved by some version of the higher-order singular value decomposition algorithm. Specifically, we simplify the STAT-SVD algorithm recently proposed by \cite{zhang2019optimal} for the tensor denoising problem, and tailor it to our generalized liquid association analysis problem to estimate $\bolGamma_{k}$, $k=1,2,3$. It consists of two major components, SVD of a matrix, and hard thresholding to identify important variables. We summarize the estimation procedure in Algorithm~\ref{algo}, then discuss each step in detail. We also note that, in our formulation, we allow the number of variables $p_k$, $k=1,2,3$, to be much larger than the sample size $n$. 

We first introduce some notation. For an integer $p$, let $[p]$ denote the set $\{1,\ldots, p\}$. For a matrix $\mbA \in \mbbR^{p \times q}$ and index sets $I\subseteq [p]$, $J\subseteq [q]$, let $\mbA_{[I,J]}$ denote the corresponding $|I|\times |J|$ submatrix, while the whole index set $[p]$ is simplified as ``:''; e.g., $\mbA_{[[p],J]} = \mbA_{[:,J]}$. Let $\mathrm{SVD}(\mbA)\in\mbbR^{p\times r}$ denote the left-$r$ singular vectors of $\mbA$, with $r \le q$. Let $\tilbolDelta_k$ and $\bolDelta_k$ denote the mode-$k$ matricization of the tensors $\bolDelta$ and $\tilbolDelta$, $k=1,2,3$. Define $\bolGamma_{-1} = \bolGamma_2 \otimes \bolGamma_3, \bolGamma_{-2} = \bolGamma_3 \otimes \bolGamma_1, \bolGamma_{-3} = \bolGamma_1 \otimes \bolGamma_2$, where $\otimes$ is the Kronecker product. Next, we define the active sets of variables in the context of generalized liquid association analysis as, 
\begin{equation} \label{activeDEF}
I_k = \left\{ j: (\bolDelta_k)_{[j,:]} \neq 0,\ 1\leq j\leq p_k \right\} \subseteq [p_k], \quad k=1,2,3.
\end{equation}
As an example, the $j$th variable $X_j$ in $\mbX$ corresponds to the $j$th row of $\bolGamma_1\in\mbbR^{p_1\times r_1}$, $j = 1, \ldots, p_1$. Therefore, variable selection in $\mbX$ translates to the row-wise sparsity in $\bolGamma_1$, and correspondingly, the row-wise sparsity in $\bolDelta_1 \in \mbbR^{p_1 \times p_2p_3}$. Define the diagonal matrix $\mbD_{I_k} \in \mbbR^{p_k \times p_k}$ that has one on the $i$th diagonal element if $i \in I_k$ and zero elsewhere. This matrix represents variable selection along each mode, and is used repeatedly in our estimation algorithm. Define $\mbD_{I_{-1}} = \mbD_{I_2} \otimes \mbD_{I_3}$, whereas $I_{-1}$ denotes the pair of subsets $I_2$ and $I_3$. Define $\mbD_{I_{-2}}, \mbD_{I_{-3}}, I_{-2}, I_{-3}$ similarly. Also define $\hatbolGamma_{-k}, \widehat{I}_k$, $k=1,2,3$, in a similar fashion.

\begin{algorithm}[t!]
\caption{generalized liquid association analysis via sparse tensor decomposition.}
\label{algo}
\begin{algorithmic}
\STATE \textbf{Input}: The centered data $\{ \mbX_i \in \mbbR^{p_1}, \mbY_i \in \mbbR^{p_2}, \mbZ_i \in \mbbR^{p_3}, i=1,\ldots,n\}$, the Tucker ranks $r_k \leq p_k$, and the sparsity parameters $(\eta_k, \widetilde{\eta}_k)$, $k=1,2,3$.

\STATE \textbf{Step 1, initialization}: Compute the sample estimate $\tilbolDelta = n^{-1} \sum_{i=1}^{n}\mbX_{i}\circ\mbY_{i}\circ\mbZ_{i}$. Obtain the initial active set $\widehat I_k^{(0)}$, and the initial basis matrices by,
\begin{equation*}
\widehat{I}_k^{(0)} = \left\{ j: \Vert(\widetilde\bolDelta_k)_{[j,:]}\Vert_{\max}> \eta_k \right\}, \quad \hatbolGamma_k^{(0)} = \mathrm{SVD}\left\{ \mbD_{\widehat{I}_k^{(0)}}\widetilde{\bolDelta}_k \mbD_{\widehat{I}_{-k}^{(0)}} \right\}\in \mbbR^{p_{k} \times r_k}, \quad k=1,2,3.
\end{equation*}

\REPEAT 
\STATE \textbf{Step 2a}: Update the active set: $\widehat{I}_k^{(t)} = \left\{ j: \|(\widetilde{\bolDelta}_k)_{[j,:]} \hatbolGamma_{-k}^{(t)}\|_2^2 > \widetilde{\eta}_k \right\}$, $k=1,2,3$.
			
\STATE \textbf{Step 2b}: Perform SVD: $\hatbolGamma_k^{(t)} = \mathrm{SVD}\left\{ \mbD_{\widehat{I}_k^{(t)}} \widetilde{\bolDelta}_k \hatbolGamma_{-k}^{(t)} \right\} \in \mbbR^{p_{k} \times r_k}$, $k=1,2,3$.
\UNTIL{some stopping criterion is met.}

\STATE \textbf{Output}: The estimated basis matrices $\hatbolGamma_k$, $k=1,2,3$, and $\hatbolDelta = \widetilde{\bolDelta} \times_1 \mbP_{\hatbolGamma_1} \times_2 \mbP_{\hatbolGamma_2} \times_3 \mbP_{\hatbolGamma_3}$.
\end{algorithmic}
\end{algorithm}

We start the algorithm by computing the sample estimate $\tilbolDelta$, then perform the initial selection of important variables and initial SVD in Step 1 of Algorithm \ref{algo}. From \eqref{activeDEF},  we see that the selection of important variables can be achieved based on $\Vert(\bolDelta_k)_{[j,:]}\Vert$ for some appropriate norm $\Vert\cdot\Vert$. In the initialization step, we employ the maximum norm, and achieve the selection by hard thresholding under the sparsity parameter $\eta_k$. The two diagonal matrices $\mbD_{\widehat{I}_{k}^{(0)}}$ and $\mbD_{\widehat{I}_{-k}^{(0)}}$ operate as the subset selection operator within the SVD operator. Moreover, depending on the sparsity parameter $\eta_k$, we may keep all the variables in the active set, i.e., $\widehat{I}_k = [p_k]$. 

Next, we iterate the algorithm by repeatedly selecting important variables and performing SVD in Step 2 of Algorithm \ref{algo}. We continue to do the selection by hard thresholding, but we use a different norm, i.e., the $\ell_2$-norm rather than the $\max$-norm, and a different sparsity parameter $\widetilde\eta_k$. This change of the norm is practically useful because of the following consideration. In the initialization step, the column dimension of $\tilbolDelta_k$ is $\prod_{k' \neq k} p_{k'}$ and is often very large, and thus the maximum norm is more effective in screening out the zero rows in $\tilbolDelta_k$. On the other hand, during the iterations, the active variable set $I_k$ is selected based on $\tilbolDelta_k\bolGamma_{-k}$, which now has a much smaller column dimension, which is $\prod_{k' \neq k} r_{k'}$. As such, the $\ell_2$-norm is preferred to being able to pick up potentially weaker signals and to refine the selection from the initialization. Moreover, a different thresholding parameter $\widetilde\eta_k$ during the iterations gives more flexibility. 

We alternate Steps 2a and 2b until some termination criterion is met. That is, we terminate the algorithm if the consecutive estimates are close enough, in that the difference between the $\ell_2$-norm of the singular values of the two iterations is smaller than $10^{-6}$, or the algorithm reaches the maximum number of iterations pre-specified at 100. In our numerical experiments, we find that the algorithm converges fast, usually within 10 to 20 iterations. We output the estimated basis matrices $\hatbolGamma_k$, $k=1,2,3$, along with the updated estimate $\hatbolDelta = \widetilde{\bolDelta} \times_1 \mbP_{\hatbolGamma_1} \times_2 \mbP_{\hatbolGamma_2} \times_3 \mbP_{\hatbolGamma_3}$ that follows a Tucker decomposition.

Our algorithm is related to the STAT-SVD algorithm of \cite{zhang2019optimal}, in that we all use hard thresholding SVD iteratively. On the other hand, \citet{zhang2019optimal} targeted a tensor denoising problem involving identically distributed normal errors, and used a double thresholding scheme with a theoretical thresholding value. Their algorithm, after obtaining the variance of the errors, became tuning-free in terms of the thresholding parameter. By contrast, we aim to obtain a low-rank tensor estimator in the context of generalized liquid association analysis. The sample estimator does not have i.i.d.\ entries, and we use a single thresholding scheme with two data-driven tuning parameters. This leads to a more flexible tuning,  and consequently an utterly different approach for the asymptotic analysis.

The thresholding values $\eta_k$ and $\widetilde\eta_k$ are treated as tuning parameters, and we propose a prediction-based approach for tuning. That is, we first randomly split the data into a training set and a testing set, and obtain the sample estimates $\tilbolDelta^\mathrm{train}$ and $\tilbolDelta^\mathrm{test}$ separately. We then apply Algorithm \ref{algo} to $\tilbolDelta^\mathrm{train}$ to obtain $\hatbolGamma_k^{\mathrm{train}}(\eta)$, $k=1,2,3$, under a given set of tuning parameters $\eta = (\eta_1,\eta_2,\eta_3, \widetilde{\eta}_1, \widetilde{\eta}_2, \widetilde{\eta}_3)$. We choose $\eta$ such that the following discrepancy is minimized, 
\begin{equation}\label{tuningCV}
L(\eta) = \left\Vert \tilbolDelta^{\mathrm{test}} - \tilbolDelta^{\mathrm{test}}\times_1 \mbP_{\hatbolGamma_1^{\mathrm{train}}(\eta)} \times_2 \mbP_{\hatbolGamma_2^{\mathrm{train}}(\eta)} \times_3 \mbP_{\hatbolGamma_3^{\mathrm{train}}(\eta)} \right\Vert_F.
\end{equation}
Meanwhile, the ranks $(r_1, r_2, r_3)$ take some pre-specified values. In practice, $r_k$ is often taken as 1 or 2 for exploratory analysis and data visualization. This is the same in spirt as canonical correlation analysis. Actually, rank selection is still an open question in both CCA and matrix or tensor denoising, and we leave a full treatment of rank selection as future research.

%%%%%%%%%%%%%%%%%%%%%%%%%%%%%%%%%%%%%%%%%%%%%%%%%%%
\section{Theoretical Properties}
\label{sec:theory}

We establish the theoretical guarantees for the estimated subspace basis matrices $\hatbolGamma_k$, the estimated Tucker tensor $\hatbolDelta = \widetilde{\bolDelta} \times_1 \mbP_{\hatbolGamma_1} \times_2 \mbP_{\hatbolGamma_2} \times_3 \mbP_{\hatbolGamma_3}$, and the estimated active sets $\widehat{I}_k^{(t)}$, $k = 1,2,3$, from Algorithm \ref{algo}. We first lay out the required regularity conditions, then derive the non-asymptotic error bounds and the asymptotic consistency. In our theoretical analysis, we allow both the tensor dimension $p = \prod_{k=1}^3 p_k$ and the sparsity level $s = \prod_{k=1}^3 s_k$ to diverge with the sample size $n$, while we fix the tensor rank $r = \prod_{k=1}^3 r_k$. Throughout this section, we use $C$ and $C^{\prime}$ to denote some generic positive constants that could vary in different contexts. 

We begin with three technical assumptions.

\vspace*{-5pt}
\begin{enumerate}[label={(A\arabic*)}]
\item \label{asm:dist} 
(Marginal distribution). Assume $\vert X_iY_j \vert \le C$ for some constant $C>0$, $i=1,\dots,p_1, j=1,\dots,p_2$. Assume $Z_k$ is sub-Gaussian with a {constant parameter $\sigma^2>0$}, $k=1,\dots,p_3$. 

\item \label{asm:singular}
(Singular value). Let $\lambda_k$ denote the smallest non-zero singular value of $\bolDelta_k$, $k=1,2,3$, and $\lambda = \min\{\lambda_1, \lambda_2, \lambda_3\}$. Assume $\lambda > \max\{C\sqrt{ s\log p/n}, C'\}$ for some constants $C, C' > 0$. 

\item \label{asm:delta}
(Signal strength). Let $\delta_{\min} = \min\limits_{k\in\{1,2,3\}, i \in I_k} \|(\bolDelta_k)_{[i,:]}\|_2$ denote the minimal signal strength. Assume $\sqrt{s\log p/n} = o(\delta_{\min})$.
\end{enumerate}
\vspace*{-5pt}

\noindent
We view these assumptions generally mild and reasonable. Particularly, Assumption \ref{asm:dist} requires $\mbX$ and $\mbY$ to be bounded and $\mbZ$ to be sub-Gaussian, which are necessary to establish the concentration of each element in $\widetilde\bolDelta$ to its population counterpart. The sub-Gaussian assumption is weaker than the normality assumption, and is widely used in high-dimensional non-asymptotic analysis \cite[see, e.g.,][]{rudelson2010non,wainwright2019high}. Besides, it assumes each individual $Z_k$ to be sub-Gaussian, which is weaker than assuming the joint distribution $\mbZ$ is sub-Gaussian. Moreover, the constant $\sigma^2>0$ does not require all $Z_k$ to have the same variance. For instance, if $Z_k$ is sub-Gaussian with parameter $\sigma_k^2$, $k=1,\dots,p_3$, then we can let $\sigma^2 = \max_k \sigma_k^2$ to have Assumption \ref{asm:dist} satisfied. Assumption \ref{asm:singular} ensures that there is a reasonable gap between the zero and nonzero eigenvalues in $\bolDelta_k$, under which the consistency for the estimator $\hatbolGamma_k$ is ensured. This type of assumption on the eigenvalues is frequently used in high-dimensional singular value decomposition literature \citep{yu2015useful, yang2016rate, zhang2019optimal}. Assumption \ref{asm:delta} guarantees that the signal of the important variables is of a reasonable magnitude when $n, p, s$ diverge, which in turn ensures the selection consistency. Note that $\delta_{\min}$ is also the minimal Frobenius norm of the non-zero slices in $\bolDelta$, i.e.,\ slices of $\bolDelta$ corresponding to those variables $i \in I_k$, $k=1,2,3$. This assumption is very mild. If we assume the nonzero entries of $\bolDelta$ are bounded away from zero, then this assumption is satisfied. 

Next, we derive the non-asymptotic error bounds of the generalized liquid association analysis estimators. Let $\hatbolDelta$ and $\hatbolGamma_k$, $k=1,2,3$, denote the estimators returned from Algorithm~\ref{algo}, using the theoretical thresholding values $\eta_k = \sqrt{\alpha \log p /n}$ and $\widetilde{\eta}_k = \alpha s_{-k} \log p/n$, where $\alpha > 512(C+\sigma)^4$ is a sufficiently large constant, and $C, \sigma$ are as defined in Assumption \ref{asm:dist}. Moreover, since the basis matrix $\hatbolGamma_k$ is identifiable only up to an orthogonal rotation, we characterize its convergence in terms of the corresponding projection matrix $\mbP_{\hatbolGamma_k}$, $k=1,2,3$. 
	
\begin{thm}[Non-asymptotic error bound] \label{thm}
Suppose Assumptions~\ref{asm:dist}  and \ref{asm:singular} hold. Then, with probability at least $1 - C \max\left[ p^{-\gamma}, p^{-\left\{ \sqrt{n(\gamma+1)/(2\log p)} - 1 \right\}} \right]$, for some constant $\gamma >0$,
\begin{enumerate}[label={(\alph*)}]
\item $\|\hatbolDelta - \bolDelta\|_F \le C\sqrt{s\log p/n}$;
\item $\max_{k=1,2,3}\| \mbP_{\hatbolGamma_k} - \mbP_{\bolGamma_k}\|_F \le C\sqrt{s\log p/n}$.
\end{enumerate}
\end{thm}

\noindent 
When the sample size is sufficiently large, i.e., when ${n} \gg 2(\gamma + 1)\log p$, both statements in Theorem~\ref{thm}  hold with probability $1 - Cp^{-\gamma}$. This probability goes to one as $p$ diverges with $n$. We also briefly comment that, although we establish Theorem~\ref{thm} for an order-3 tensor, the results can be straightforwardly extended for a general order tensor. 

Next, we establish the asymptotic consistency as $n, p, s$ diverge to infinity. 
\begin{thm}[Asymptotic consistency] \label{thm:asym}
Suppose Assumptions~\ref{asm:dist} to \ref{asm:delta} hold. Then, with probability tending to one, as $n, p, s \to \infty$, we have,
\begin{enumerate}[label={(\alph*)}]
\item $\|\hatbolDelta - \bolDelta\|_F \rightarrow 0$;
\item $\| \mbP_{\hatbolGamma_k} - \mbP_{\bolGamma_k}\|_F  \rightarrow 0$, $k = 1,2,3$;
\item $\widehat{I}_k^{(t)} = I_k$, $k = 1,2,3$, and for iterations $t = 0,1,\ldots, t_{\max}$.
\end{enumerate}
\end{thm}

\noindent
Theorem~\ref{thm:asym} establishes the consistency of tensor parameter estimation, subspace estimation, as well as variable selection, under $s\log p = o(n)$, which allows each tensor dimension $p_k$ to diverge faster than the growing sample size $n$. Moreover, the consistent variable selection is established for every iteration of Algorithm \ref{algo}. Note that the maximum number of iterations $t_{\max}$ in Theorem~\ref{thm:asym} (c) is treated as a constant regardless of the tensor dimension or sample size. In practice, we have observed that the algorithm often converges within 10 to 20 iterations. As such, we feel it reasonable to treat $t_{\max}$ as a constant in our theoretical analysis. On the other hand, we can extend the result by allowing $t_{\max}$ to diverge as well.  For instance, parallel to the arguments in \cite{zhang2019optimal}, we can let $t_{\max}$ diverge at the rate of $o(p)$. 

Theorems~\ref{thm} and~\ref{thm:asym} provide the theoretical guarantees for our proposed generalized liquid association analysis in the high dimensional setting. At the same time, these results may be of independent interest for more general tensor estimation problems.

%%%%%%%%%%%%%%%%%%%%%%%%%%%%%%%%%%%%%%%%%%%%%%%%%%%
\section{Simulation Studies}
\label{sec:sim}

%%%%%%%%%%%%%%%%%%%%%%%%%%%%%%%%%%%%%%%%%%%%%%%%%%%
\subsection{Simulation setup}

We carry out the simulations to investigate the empirical performance of the proposed generalized liquid association analysis (GLAA) method. We consider three scenarios. In the first scenario, we fix the dimension of $\mbZ$ at $p_3=1$, and increase the dimensions of $\mbX$ and $\mbY$ as $p_1 = p_2 = \{100, 200, 300, 400, 500\}$. In the second scenario, we fix $p_1 = p_2 = 100$, and increase $p_3 = \{20, 40, 60, 80, 100\}$. In both cases, we fix the sample size at $n=500$. In the third scenario, we fix $p_1 = 100, p_2 = 25, p_3 = 1$, and increase the sample size $n = \{60, 80, 100, 120, 160\}$. 

We generate the data in the following way. For $i=1,\ldots,n$, we first generate $\mbZ_i$ from a normal distribution with mean zero and covariance $\mbI_{p_3}$. We then generate $(\mbX_i, \mbY_i)$ jointly from a normal distribution with mean zero and covariance, \begin{equation*}
\Cov(\mbX, \mbY \mid \mbZ = \mbZ_i )= 
\begin{pmatrix}
\bolSigma_{\mbX} & \bolGamma_1 \mbf(\bolGamma_3^\top \mbZ_i) \bolGamma_2^\top \\
\bolGamma_2 \mbf^\top(\bolGamma_3^\top \mbZ_i) \bolGamma_1^\top & \bolSigma_{\mbY}
\end{pmatrix}.
\end{equation*}
To ensure the positive-definiteness of this covariance matrix, we set $\bolGamma_1 = \bolSigma_{\mbX}^{1/2} (\mbO_1^\top, \mbzero)^\top$ and $\bolGamma_2 = \bolSigma_{\mbY}^{1/2} (\mbO_2^\top, \mbzero)^\top$, where $\mbO_1 = \mbO_2 \in \mbbR^{5\times 2}$, with the first column being $(1,1,1,1,1,0,\ldots,0)/$ $\sqrt{5}$, and the second column being $(0,0,0,-1,1,0,\ldots,0)/\sqrt{2}$. As a result, in this example, for $\mbX$ and $\mbY$, the ranks are $r_1 = r_2 = 2$, and the sparsity levels are $s_1 = s_2 = 5$. The marginal covariance matrix $\bolSigma_{\mbX}$ is set as a block diagonal matrix, $\bolSigma_{\mbX} = \mathrm{bdiag}(\bolSigma_{\mbX,1}, \bolSigma_{\mbX,2})$, where $\bolSigma_{\mbX,1} \in\mbbR^{s_1 \times s_1}$ corresponds to the active variables in $\mbX$ and takes the form of an AR structure such that its $(i,j)$th entry equals $\sigma_{ij} = 0.3^{|i-j|}$, $i, j = 1,\ldots, s_1$, and $\bolSigma_{\mbX,2}\in\mbbR^{(p_1-s_1) \times (p_1-s_1)}$ is the identity matrix. The marginal covariance matrix $\bolSigma_{\mbY}$ is constructed in a similar fashion. The matrix $\mbf(\bolGamma_3^\top \mbZ_i)$ is set as a diagonal matrix, $\mathrm{diag}\{f_1(\bolGamma_3^\top \mbZ_i), f_2(\bolGamma_3^\top \mbZ_i)\}$, where $f_1(a) = 0.95 \mathrm{sign}(a)$ and $f_2(a) = 0.85 \mathrm{sign}(a)$. In the Supplementary Appendix, we consider additional simulations using $f(a; \rho, \xi) = \rho\{2/(1+e^{-2\xi a}) - 1\}$ with different parameters $0 < \rho \le 1$ and $\xi > 0$ that control the magnitude and speed of changes in $\Cov(\mbX, \mbY | \mbZ)$. For the first and the third scenarios, $p_3=1$ and thus $\bolGamma_3 = 1$. For the second scenario, where $p_3$ varies from $20$ to $100$, we set $\bolGamma_3 = (1,1,1,1,1,0,\ldots,0)/\sqrt{5}$, with $s_3 = 5$.  

When applying the proposed method, we choose $\eta_1$ and $\eta_2$, which are the thresholding parameters only used in the initialization step of the algorithm, so that about half of the variables in $\mbX$ and in $\mbY$ are kept for subsequent iterations. We then tune $\widetilde\eta_1$ and $\widetilde\eta_2$, the thresholding parameters used in iterative sparse SVD, by cross-validation over a grid of candidate values. We choose the ranks $r_1, r_2$, i.e., the numbers of linear combinations for $\mbX$ and $\mbY$, to be one, which is most commonly used in canonical correlation analysis as well. 

There is no existing method designed to directly address our targeting problem. For the purpose of comparison, we consider three relevant solutions. The first solution we consider is a naive and marginal extension of the univariate liquid association (ULA) method from \citet{li2002LA}. That is, we construct a tensor estimator $\tilbolPhi$, each entry of which is the sample univariate LA for the triplet $(X_{i_1}, Y_{i_2}, Z_{i_3})$ as defined in \citet{li2002LA}, $i_1=1,\ldots,p_1, i_2=1,\ldots,p_2, i_3=1,\ldots,p_3$. We then perform the usual SVD to each matricization of $\tilbolPhi$ under the given rank to obtain the estimates of basis matrices; i.e., $\mathrm{SVD}\big\{ \tilbolPhi_{(k)} \big\}$, $k=1,2,3$. The second and third solutions we consider are two different versions of canonical correlation analysis, the penalized matrix decomposition (PMD) method of \citet{witten2009penalized}, and the sparse canonical correlation analysis (SCCA) method of \citet{mai2019iterative}. We have chosen these two versions due to their computational simplicity and superior empirical performance. We note that CCA is not designed to incorporate the third set of variables $\mbZ$. We thus simply take the first $r_1$ and $r_2$ directions of $\mbX$ and $\mbY$ from CCA as the estimated basis matrices corresponding to $\bolGamma_1$ and $\bolGamma_2$. We evaluate the performance of each method in terms of the variable selection accuracy and the subspace estimation accuracy. 

For variable selection, we record the true positive rate (TPR) and false positive rate (FPR) for each mode. Recall from \eqref{activeDEF}, the active set of variables is $I_k$, which is also the index set of nonzero rows in $\bolGamma_k$. Let $\widehat{I}_k$ denote the estimated active set corresponding to $\hatbolGamma_k$, then $\textrm{TPR-}k = |I_k \cap \widehat{I}_k| / s_k$ and $\textrm{FPR-}k = |I_k^c \cap \widehat{I}_k| / (p_k - s_k)$, $k = 1,2,3$. For GLAA, PMD and SCCA, we estimate the active set as $\widehat{I}_k = \left\{ i: \text{there exist non-zero elements in the $i$th row of } \hatbolGamma_k \right\}$. For ULA, it does not perform any variable selection. For the purpose of comparison, we simply calculate the $\ell_2$-norm of each row for the $k$th matricization $\tilbolPhi_{(k)}$, arrange the row indices in a descending order by the $\ell_2$-norms, then select the first $s_k$ rows for each mode, $k=1,2,3$. Of course, the information about $s_k$ is generally unknown in practice, and this solution utilizes such knowledge. Even so, as we show later, ULA is still far less effective compared to the proposed GLAA method. 

For subspace estimation, we compute the average distance between the true and the estimated subspaces, $D = \sum_{k=1}^{\tilde k} D(\bolGamma_k, \hatbolGamma_k) / \tilde{k}$, where $D(\bolGamma_k,\hatbolGamma_k) = \big\Vert \mbP_{\bolGamma_k}-\mbP_{\hatbolGamma_k} \big\Vert_{F} /\sqrt{2r_k}$. For GLAA and ULA, this distance measure is averaged over all three modes of $\mbX, \mbY, \mbZ$, so $\tilde{k} = 3$. For PMD and SCCA, it is averaged over the first two modes $\mbX, \mbY$, so $\tilde{k} = 2$. By definition, this distance measure is between 0 and 1, where 0 indicates a perfect subspace recovery.

%%%%%%%%%%%%%%%%%%%%%%%%%%%%%%%%%%%%%%%%%%%%%%%%%%%
\subsection{Simulation results}

Tables \ref{tab:p}--\ref{tab:n} summarize the simulation results over 100 data replications for the three scenarios.

\begin{table}[t!]
\centering
\resizebox{\textwidth}{!}{
\begin{tabular}{cl|ccccc} \toprule
$p_1, p_2$ & Method & TPR-1 & FPR-1 & TPR-2 & FPR-2 & $D$\\ \hline
			\multirow{4}{*}{100} 
			& GLAA & 1.000 (0.000) & 0.000 (0.000) & 1.000 (0.000) & 0.000 (0.000) & 0.095 (0.002) \\ 
			& ULA & 0.998 (0.002) & 0.000 (0.000) & 0.996 (0.003) & 0.000 (0.000) & 0.776 (0.002)  \\ 
			& PMD & 0.804 (0.028) & 0.735 (0.029) & 0.802 (0.029) & 0.731 (0.028) & 0.971 (0.002) \\ 
			& SCCA & 0.584 (0.030) & 0.626 (0.027) & 0.634 (0.030) & 0.629 (0.027) & 0.989 (0.001) \\ \hline
			\multirow{4}{*}{200} 
			& GLAA & 1.000 (0.000) & 0.000 (0.000) & 1.000 (0.000) & 0.000 (0.000) & 0.100 (0.003)  \\ 
			& ULA  & 0.928 (0.010) & 0.002 (0.000) & 0.944 (0.009) & 0.001 (0.000) &0.873 (0.002)  \\ 
			& PMD & 0.766 (0.033) & 0.731 (0.029) & 0.762 (0.033) & 0.727 (0.029) & 0.989 (0.001)  \\ 
			& SCCA & 0.596 (0.031) & 0.611 (0.027) & 0.626 (0.035) & 0.611 (0.027) & 0.993 (0.001)  \\  \hline
			\multirow{4}{*}{300} 
			& GLAA & 1.000 (0.000) & 0.000 (0.000) & 1.000 (0.000) & 0.000 (0.000) & 0.100 (0.003) \\ 
			& ULA  & 0.808 (0.016) & 0.003 (0.000) & 0.806 (0.016) & 0.003 (0.000) & 0.945 (0.003) \\ 
			& PMD & 0.718 (0.032) & 0.693 (0.030) & 0.730 (0.034) & 0.696 (0.029) & 0.992 (0.001) \\ 
			& SCCA & 0.670 (0.028) & 0.651 (0.019) & 0.624 (0.028) & 0.651 (0.018) & 0.997 (0.000) \\ \hline
			\multirow{4}{*}{400} 
			& GLAA & 0.932 (0.024) & 0.008 (0.004) & 0.930 (0.025) & 0.008 (0.004) & 0.186 (0.025) \\ 
			& ULA  & 0.652 (0.018) & 0.004 (0.000) & 0.678 (0.019) & 0.004 (0.000) & 0.980 (0.002) \\ 
			& PMD & 0.798 (0.029) & 0.766 (0.026) & 0.800 (0.029) & 0.762 (0.026) & 0.994 (0.001) \\ 
			& SCCA & 0.594 (0.022) & 0.601 (0.010) & 0.590 (0.024) & 0.602 (0.010) & 0.997 (0.000) \\ \hline
			\multirow{4}{*}{500} 
			& GLAA & 0.848 (0.032) & 0.041 (0.010) & 0.848 (0.033) & 0.040 (0.009) & 0.304 (0.037) \\ 
			& ULA  & 0.526 (0.021) & 0.005 (0.000) & 0.526 (0.020) & 0.005 (0.000) & 0.989 (0.001) \\ 
			& PMD & 0.788 (0.031) & 0.727 (0.027) & 0.794 (0.030) & 0.729 (0.028) & 0.995 (0.001) \\ 
			& SCCA & 0.532 (0.024) & 0.518 (0.008) & 0.532 (0.024) & 0.516 (0.008) & 0.997 (0.000) \\ \bottomrule
\end{tabular}
}
\caption{The variable selection accuracy, measured by TPR and FPR, and the subspace estimation accuracy, measured by $D$, for Scenario 1 where $p_1=p_2$ varies. The reported are the average criteria, with the standard errors in the parenthesis, over 100 data replications.} \label{tab:p}
\end{table}

\begin{table}[t!]
\centering
\resizebox{\textwidth}{!}{
\begin{tabular}{cl|cccccc} \toprule
$p_3$ & Method & TPR-1 & FPR-1 & TPR-2 & FPR-2 & $D$ & \\ \hline
			\multirow{4}{*}{20} 
			& GLAA & 1.000 (0.000) & 0.000 (0.000) & 1.000 (0.000) & 0.000 (0.000) & 0.132 (0.004) \\ 
			& ULA  & 0.642 (0.019) & 0.019 (0.001) & 0.638 (0.019) & 0.019 ( 0.001) & 0.872 (0.003) \\ 
			& PMD & 0.774 (0.031) & 0.733 (0.029) & 0.776 (0.031) & 0.732 (0.029) & 0.978 (0.002)  \\ 
			& SCCA & 0.518 (0.034) & 0.534 (0.029) & 0.536 (0.032) & 0.532 (0.030) & 0.989 (0.001) \\ \hline
			\multirow{4}{*}{40} 
			& GLAA & 1.000 (0.000) & 0.000 (0.000) &1.000 (0.000) & 0.000 (0.000) & 0.131 (0.004) \\ 
			& ULA  & 0.488 (0.020) & 0.027 (0.001) & 0.498 (0.020) & 0.026 (0.001) & 0.888 (0.002) \\ 
			& PMD & 0.774 (0.031) & 0.695 (0.030) & 0.772 (0.032) & 0.695 (0.029) & 0.973 (0.002) \\ 
			& SCCA & 0.590 (0.034) & 0.596 (0.031) & 0.624 (0.033) & 0.589 (0.031) & 0.987 (0.001) \\ \hline
			\multirow{4}{*}{60} 
			& GLAA &1.000 (0.000) & 0.000 (0.000) & 1.000 (0.000) & 0.000 (0.000) & 0.136 (0.004) \\ 
			& ULA  & 0.384 (0.020) & 0.032 (0.001) & 0.404 (0.020) & 0.031 (0.001) & 0.898 (0.002) \\ 
			& PMD & 0.748 (0.030) & 0.694 (0.030) & 0.754 (0.033) & 0.701 (0.030) & 0.973 (0.002) \\ 
			& SCCA & 0.564 (0.031) & 0.537 (0.029) & 0.572 (0.031) & 0.541 (0.029) & 0.985 (0.002) \\ \hline
			\multirow{4}{*}{80} 
			& GLAA & 0.998 (0.002) & 0.000 (0.000) & 0.998 (0.002) & 0.000 (0.000)& 0.145 (0.005) \\ 
			& ULA  & 0.378 (0.021) & 0.033 (0.001) & 0.368 (0.019) & 0.033 (0.001) & 0.903 (0.001) \\ 
			& PMD & 0.706 (0.033) & 0.689 (0.031) & 0.760 (0.032) & 0.688 (0.031) & 0.975 (0.002) \\ 
			& SCCA & 0.624 (0.031) & 0.613 (0.027) & 0.592 (0.033) & 0.609 (0.027) & 0.987 (0.002) \\ \hline
			\multirow{4}{*}{100} 
			& GLAA & 0.996 (0.003) & 0.005 (0.005) & 0.996 (0.003) & 0.005 (0.005) & 0.158 (0.010) \\ 
			& ULA  & 0.332 (0.020) & 0.035 (0.001) & 0.360 (0.018) & 0.034 (0.001) & 0.905 (0.001) \\ 
			& PMD & 0.712 (0.034) & 0.633 (0.033) & 0.662 (0.038) & 0.633 (0.033)  & 0.974 (0.002) \\ 
			& SCCA & 0.560 (0.033) & 0.565 (0.030) & 0.586 (0.032) & 0.567 (0.030) & 0.989 (0.001) \\ \bottomrule
\end{tabular}
}
\caption{The variable selection accuracy, measured by TPR and FPR, and the subspace estimation accuracy, measured by $D$, for Scenario 2 where $p_3$ varies. The reported are the average criteria, with the standard errors in the parenthesis, over 100 data replications.}
\label{tab:p3}
\end{table}

\begin{table}[t!]
\centering
\resizebox{\textwidth}{!}{
\begin{tabular}{clccccc} \toprule
$n$ & Method & TPR-1 & FPR-1 & TPR-2 & FPR-2 & $D$ \\ \hline
			\multirow{4}{*}{60} 
			& GLAA & 0.606 (0.025) & 0.048 (0.008) & 0.664 (0.027) & 0.154 (0.021) & 0.743 (0.015) \\ 
			& ULA  & 0.410 (0.022) & 0.031 (0.001) & 0.362 (0.018) & 0.160 (0.004) & 0.920 (0.004)  \\ 
			& PMD & 0.498 (0.040) & 0.486 (0.036) & 0.552 (0.039) & 0.466 (0.036) & 0.957 (0.004)  \\ 
			& SCCA & 0.604 (0.023) & 0.059 (0.012) & 0.688 (0.023) & 0.643 (0.017) & 0.968 (0.002)  \\ \hline
			\multirow{4}{*}{80} 
			& GLAA & 0.638 (0.027) & 0.018 (0.002) & 0.738 (0.025) & 0.137 (0.027) & 0.662 (0.021) \\ 
			& ULA  & 0.550 (0.021) & 0.024 (0.001) & 0.412 (0.018) & 0.147 (0.005) & 0.895 (0.004)  \\ 
			& PMD & 0.494 (0.038) & 0.441 (0.034) & 0.502 (0.038) & 0.429 (0.034) & 0.953 (0.004)  \\ 
			& SCCA & 0.672 (0.026) & 0.665 (0.016) & 0.716 (0.025) & 0.718 (0.019) & 0.966 (0.003)  \\ \hline
			\multirow{4}{*}{100} 
			& GLAA & 0.820 (0.022) & 0.013 (0.004) & 0.848 (0.020) & 0.060 (0.016) & 0.483 (0.024) \\ 
			& ULA  & 0.686 (0.019) & 0.017 (0.001) & 0.534 (0.021) & 0.117 (0.005) & 0.870 (0.004)  \\ 
			& PMD & 0.510 (0.035) & 0.476 (0.032) & 0.560 (0.036) & 0.442 (0.033) & 0.947 (0.004)  \\ 
			& SCCA & 0.702 (0.027) & 0.677 (0.021) & 0.732 (0.025) & 0.714 (0.024) & 0.961 (0.003)  \\ \hline
			\multirow{4}{*}{120} 
			& GLAA & 0.896 (0.018) & 0.004 (0.001) & 0.914 (0.018) & 0.010 (0.003) & 0.350 (0.021)  \\ 
			& ULA  & 0.802 (0.017) & 0.010 (0.001) & 0.622 (0.018) & 0.094 (0.004) & 0.850 (0.003)  \\ 
			& PMD & 0.576 (0.035) & 0.564 (0.032) & 0.662 (0.035) & 0.530 (0.032) & 0.941 (0.004)  \\ 
			& SCCA & 0.640 (0.030) & 0.657 (0.024) & 0.708 (0.028) & 0.705 (0.022) & 0.970 (0.003)  \\ \hline
			\multirow{4}{*}{160} 
			& GLAA & 0.990 (0.004) & 0.001 (0.000) & 0.984 (0.005) & 0.002 (0.001) & 0.209 (0.012) \\ 
			& ULA  & 0.920 (0.011) & 0.004 (0.001) & 0.732 (0.016) & 0.067 (0.004) & 0.812 (0.003)  \\ 
			& PMD & 0.610 (0.037) & 0.570 (0.034) & 0.652 (0.038) & 0.544 (0.035) & 0.944 (0.004)   \\ 
			& SCCA & 0.626 (0.031) & 0.651 (0.024) & 0.698 (0.029) & 0.0691 (0.024) & 0.969 (0.003)  \\ \bottomrule
\end{tabular}
}
\caption{The variable selection accuracy, measured by TPR and FPR, and the subspace estimation accuracy, measured by $D$, for Scenario 3 where $n$ varies. The reported are the average criteria, with the standard errors in the parenthesis, over 100 data replications.}
\label{tab:n}
\end{table}

Table \ref{tab:p} reports the accuracy of variable selection and subspace estimation when the dimension $p_1=p_2$ of $\mbX$ and $\mbY$ increases. It is clearly seen that GLAA dominates all the competing solutions. For ULA, variable selection and subspace estimation are carried out separately. Even with the oracle knowledge of the true sparsity level, the naive variable selection of ULA still performs worse, as it only utilizes the marginal information of each mode. Besides, the estimated subspace is distant away from the true subspace, especially when $p_1, p_2$ are large. Moreover, we see that, as $p_1, p_2$ increase, the performance of ULA degrades fast, while GLAA remains competitive. For PMD and SCCA, both suffer large false positive rates in selection, while the estimated subspaces are almost orthogonal to the true subspace, as reflected by the distance measure $D$ that is almost one. This is due to the fact that, by design, neither method takes into account the conditioning variable $\mbZ$ information when studying the association between $\mbX$ and $\mbY$.  

Table \ref{tab:p3} reports the variable selection and subspace estimation accuracy when the dimension $p_3$ of $\mbZ$ increases. In this case, our goal is to estimate the subspaces spanned by $\bolGamma_1, \bolGamma_2, \bolGamma_3$ accurately, meanwhile select the variables in $\mbX$ and $\mbY$ accurately. It is seen that GLAA again outperforms all other methods considerably. Besides, it shows a competitive performance of GLAA even with a relatively large dimensionality of $\mbZ$. This also complements our real data example where the dimension of $\mbZ$ is one. 

Table \ref{tab:n} reports the variable selection and subspace estimation accuracy when the sample size $n$ increases. The size of $n$ we examine is comparable to that in our multimodal PET example. It is seen that GLAA performs the best, even under a relatively small sample size. Moreover, the performances of all methods improve as $n$ increases. However, ULA suffers a poor subspace estimation accuracy, while both PMD and SCCA continue to suffer both high false positive rates and poor subspace estimation accuracy, even for a relatively large sample size.

%%%%%%%%%%%%%%%%%%%%%%%%%%%%%%%%%%%%%%%%%%%%%%%%%%%
\section{Multimodal PET Analysis}
\label{sec:realdata}

%%%%%%%%%%%%%%%%%%%%%%%%%%%%%%%%%%%%%%%%%%%%%%%%%%%
\subsection{Study and data description}

We revisit the multimodal PET study introduced in Section \ref{subsec:motivation}. It is part of the ongoing Berkeley Aging Cohort Study that targets Alzheimer's disease (AD) as well as normal aging. AD is an irreversible neurodegenerative disorder and the leading form of dementia. It is characterized by progressive impairment of cognitive and memory functions, then loss of bodily functions, and ultimately death. AD and related dementia currently affects more than 10\% of adults aged 65 or older, and the prevalence is continuously growing. It has now become an international imperative to understand, diagnose, and treat this disorder \citep{AD2020}. 

The data consist of $n=81$ elderly subjects, with the average age $77.5$ years, and the standard deviation $6.2$ years. For each subject, three types of neuroimages were acquired, including a Pittsburgh Compound B (PiB) PET scan that measures amyloid-beta protein, an AV-1451 PET scan that measures tau protein, and a 1.5T structural MRI scan that was used for coregistration. MRI and PET images have all been preprocessed, and PET images were both coregistered to each participant's MRI image. Moreover, a mask representing voxels likely to accumulate cortical amyloid and tau pathology was created, by intersecting a cortical brain mask from the Automated Anatomical Labeling atlas \citep{TzourioMazoyer2002}, with a mask of high-probability grey matter voxels from the SPM12 tissue probability map \citep{Lockhart2017}. Then a set of MNI-space regions of interest were created, and the amount of amyloid-beta and tau deposition was summarized for each region. This results in $p_1=60$ regions for amyloid-beta PET, and $p_2=26$ regions for tau-PET. We note that brain region parcellation is particularly  useful to facilitate the interpretation, and has been frequently employed in brain imaging analysis \citep{Fornito2013, Kang2016}.

%%%%%%%%%%%%%%%%%%%%%%%%%%%%%%%%%%%%%%%%%%%%%%%%%%%
\subsection{Analyses and results}

One of the primary goals of this study is to identify brain regions where the association of amyloid-beta and tau changes the most as age varies, and to further understand this association change. We cast this problem in the framework of liquid association analysis. Let $\mbX \in \mbbR^{60}$, $\mbY \in \mbbR^{26}$ denote the amyloid-beta  accumulation and tau accumulation in various brain regions, respectively, and $Z \in \mbbR$ denote the subject's age. We first log-transform each variable in $\mbX$ and $\mbY$, and standardize $\mbX$, $\mbY$ and $Z$ marginally. We then apply the proposed generalized liquid association analysis (GLAA) method to this data. 

\begin{figure}[t!]
\centering
\begin{minipage}{0.495\textwidth}
\includegraphics[width=\linewidth]{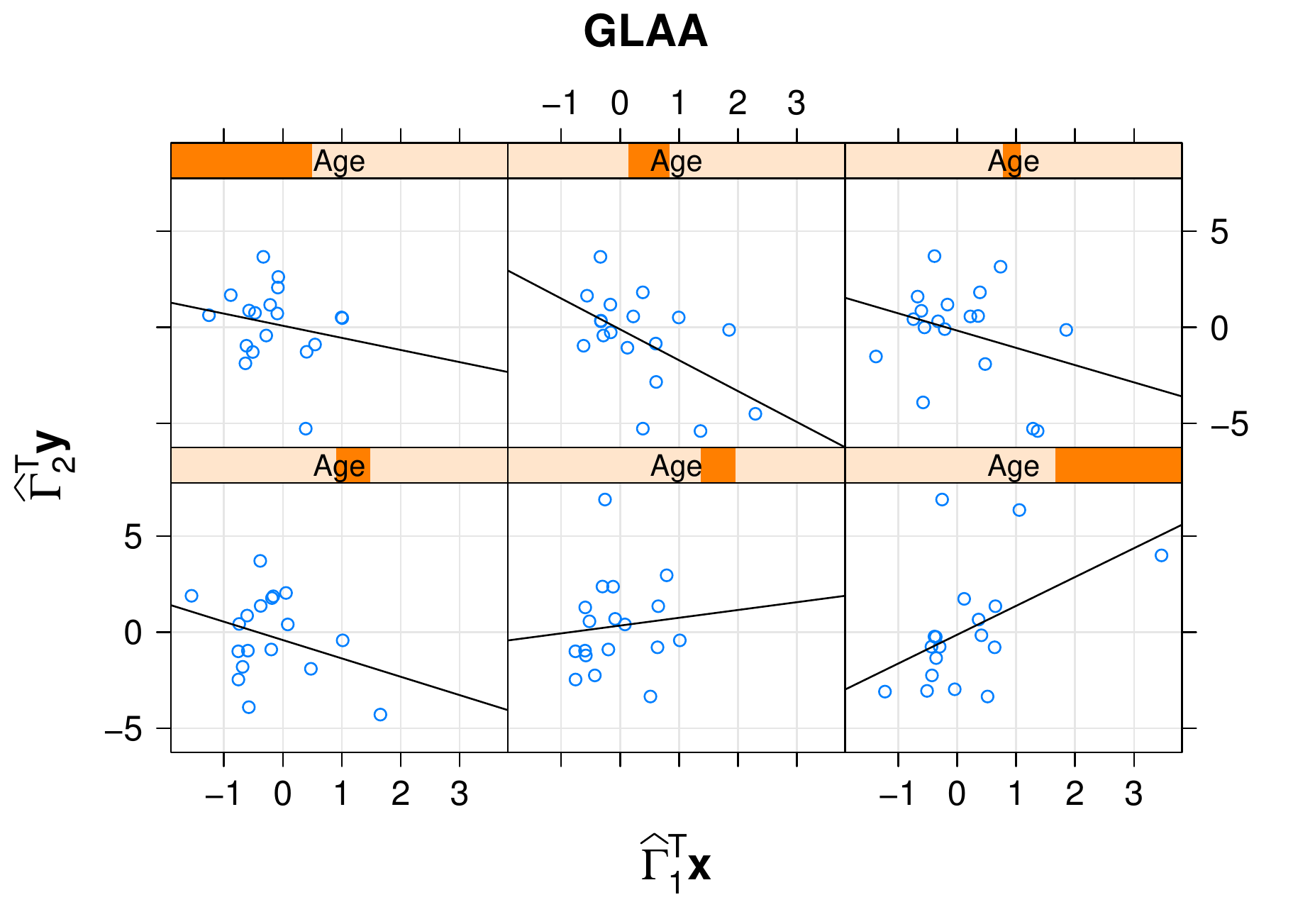}
\end{minipage}
\begin{minipage}{0.495\textwidth}
\includegraphics[width=\linewidth]{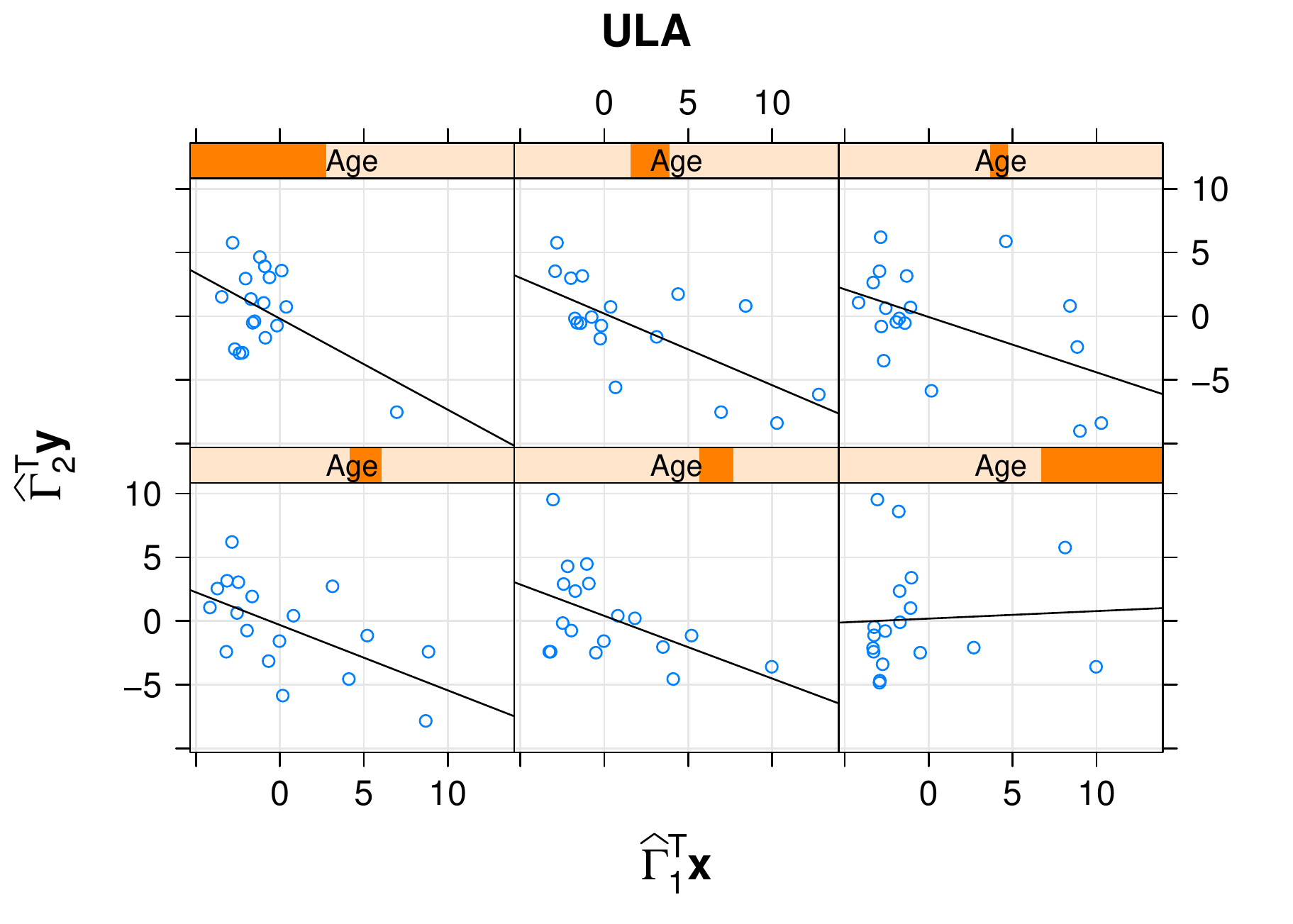}
\end{minipage}
\begin{minipage}{0.495\textwidth}
\includegraphics[width=\linewidth]{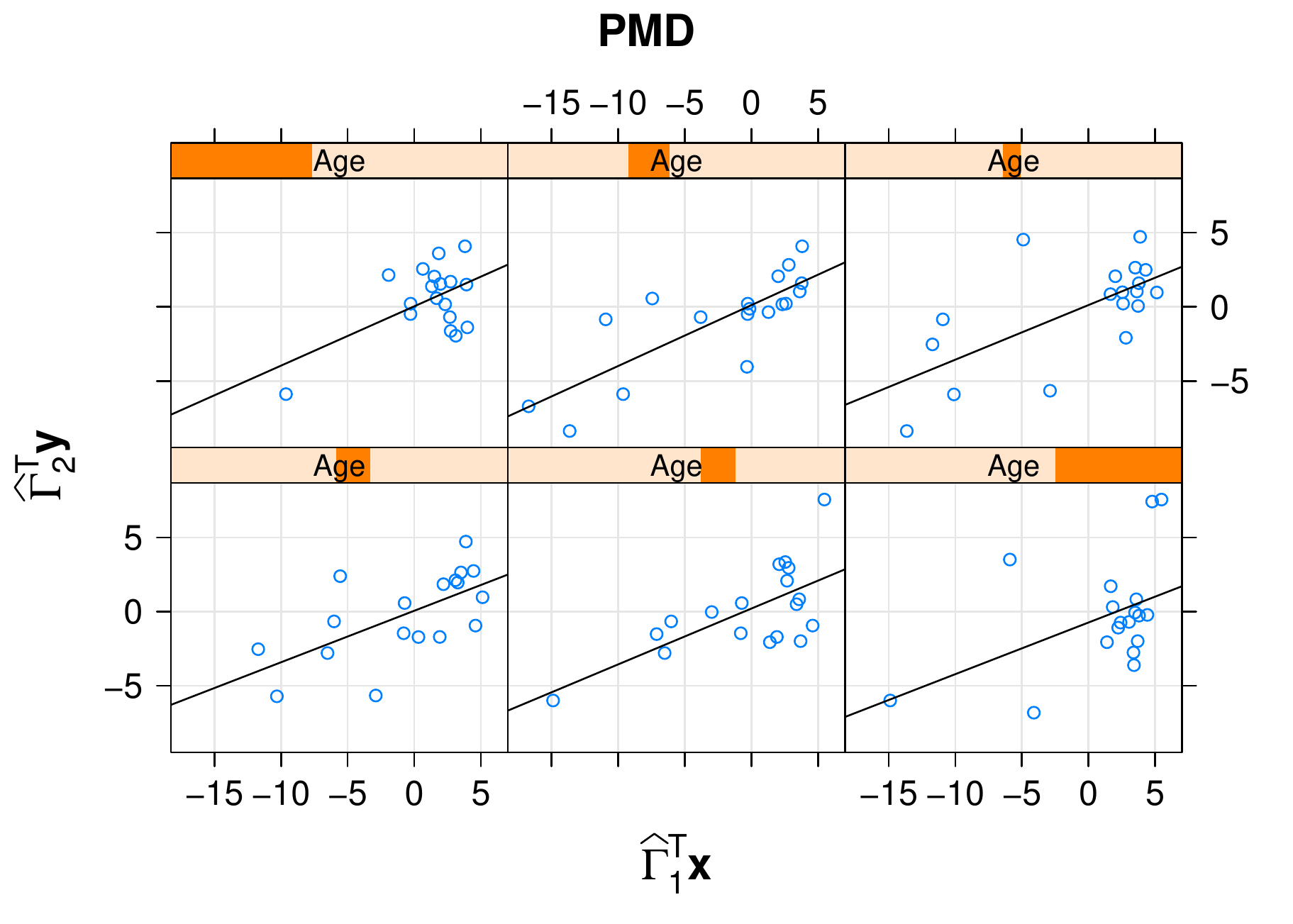}
\end{minipage}
\begin{minipage}{0.495\textwidth}
\includegraphics[width=\linewidth]{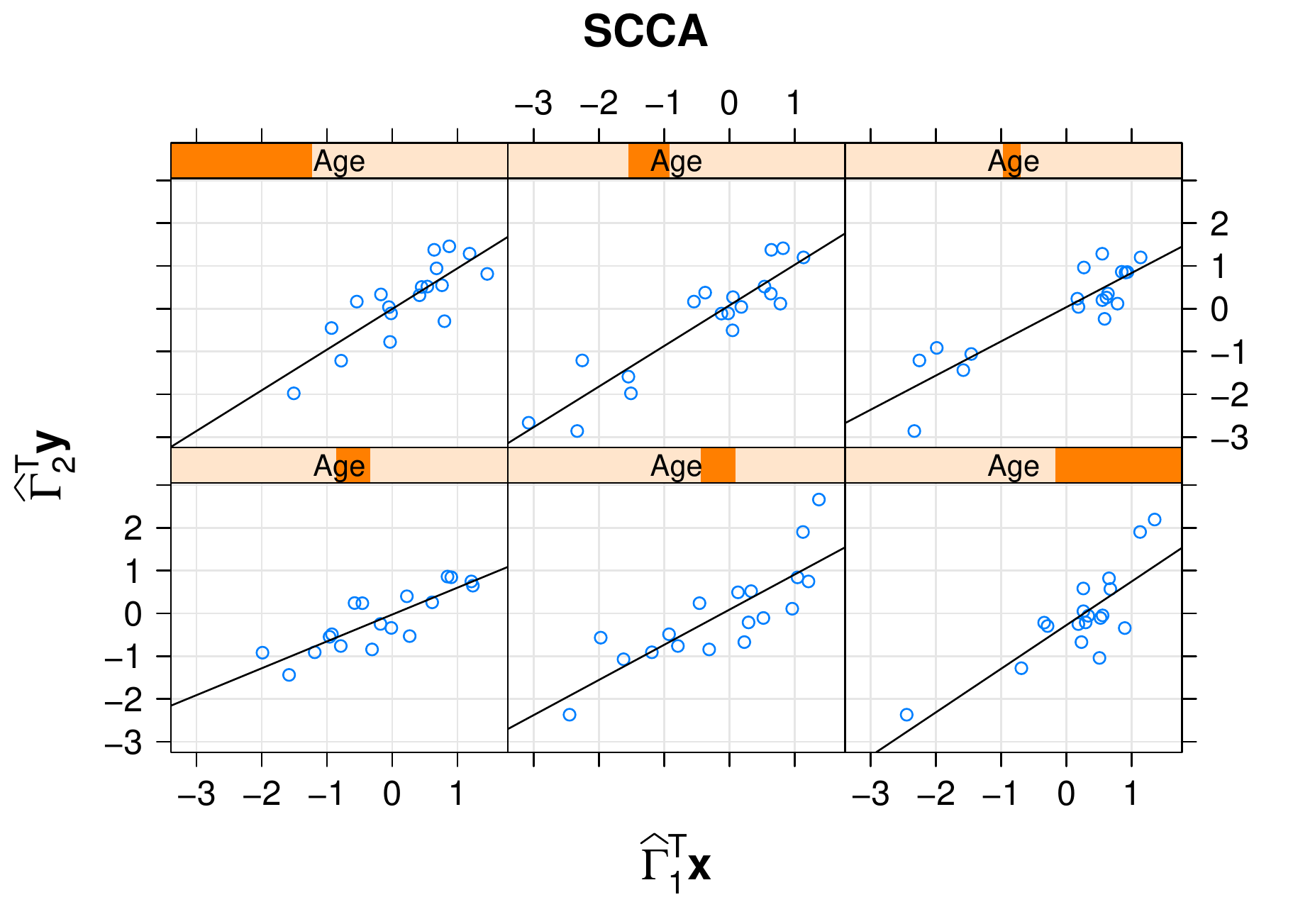}
\end{minipage}
\caption{Trellis plots of the estimated linear combinations $\hatbolGamma_2^\top \mbY$ versus $\hatbolGamma_1^\top \mbX$ as $\mbZ$ varies. Each panel represents an interval of $\mbZ$, with a linear regression line added. The methods under comparison are: generalized liquid association analysis (GLAA), univariate liquid association (ULA), penalized matrix decomposition (PMD), and sparse canonical correlation analysis (SCCA).}
\label{fig:age}
\end{figure}

After obtaining the two estimated linear combinations $\hatbolGamma_1^\top \mbX$ and $\hatbolGamma_2^\top \mbY$, we plot them as the value of $Z$ changes. We divide the interval of $Z$ into six equal-sized intervals with overlaps, then draw the scatterplot of $\hatbolGamma_2^\top \mbY$ versus $\hatbolGamma_1^\top \mbX$ within each interval. We also add a fitted linear regression line in each panel to reflect the correlation between $\hatbolGamma_1^\top \mbX$ and $\hatbolGamma_2^\top \mbY$. Figure~\ref{fig:age} shows the trellis plots, where the stripe at the top of each panel represents the range of $Z$ it covers. It is interesting to see from the GLAA estimation, the correlation between $\bolGamma_1^\top \mbX$ and $\bolGamma_2^\top \mbY$ changes from negative to positive gradually, as the age variable $Z$ increases. This may be due to different deposition patterns of amyloid-beta and tau. In particular, amyloid-beta plaques are detectable in the brain many years before dementia onset, while tau neurofibrillary tangles aggregate specifically in the medial temporal lobes in normal aging. The spread of tau out of medial temporal lobes and into the surrounding isocortex at elder age coincides with cognitive impairment, and the process is hypothesized to be potentiated or accelerated by the presence of amyloid-beta \citep{He2018, Vogel2020}. The change from a negative association in early years to a positive association in later years between amyloid-beta and tau found by our GLAA method may offer some support to this hypothesis. As a comparison, no clear changing pattern is observed from the other three estimation methods.

\begin{figure}[t!]
\centering
\begin{tabular}{cc}
\includegraphics[width=3.1in,height=2in]{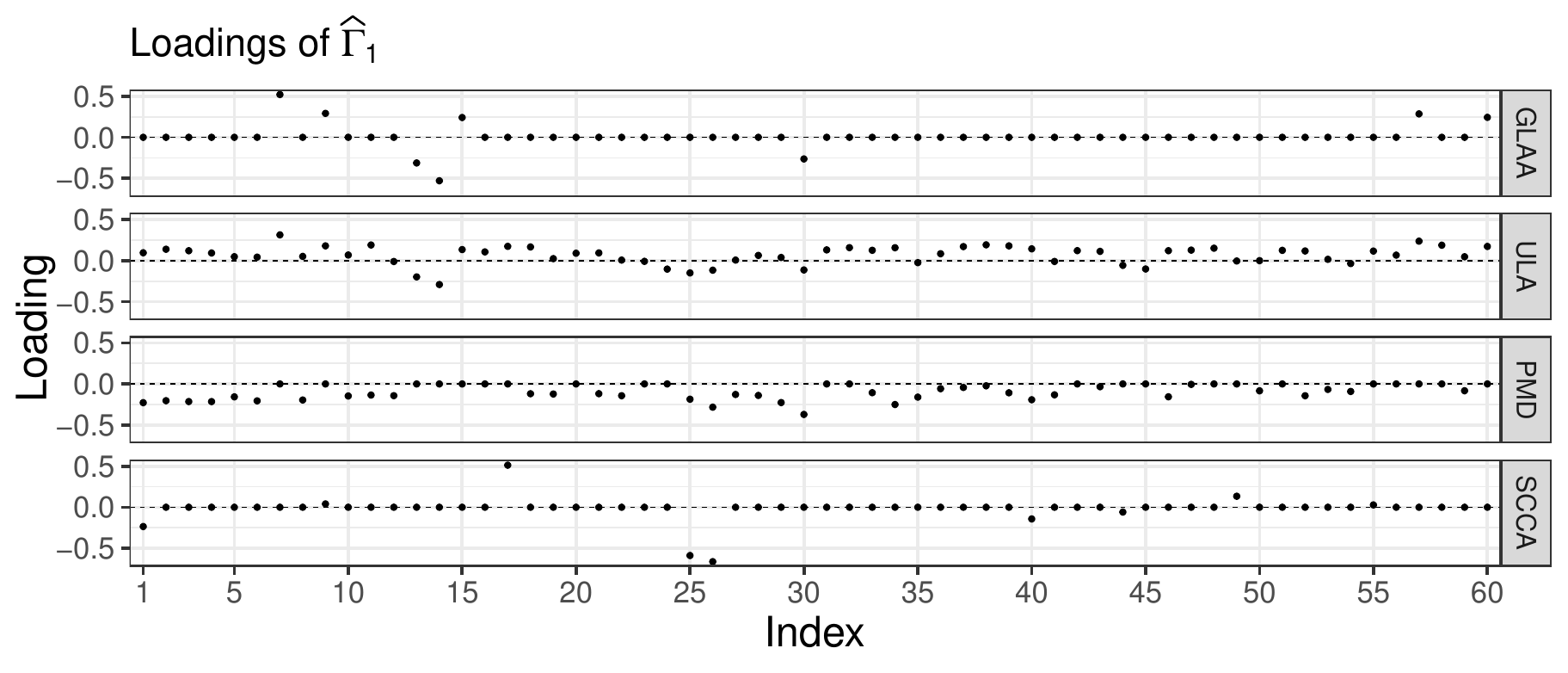} &
\includegraphics[width=3.1in,height=2in]{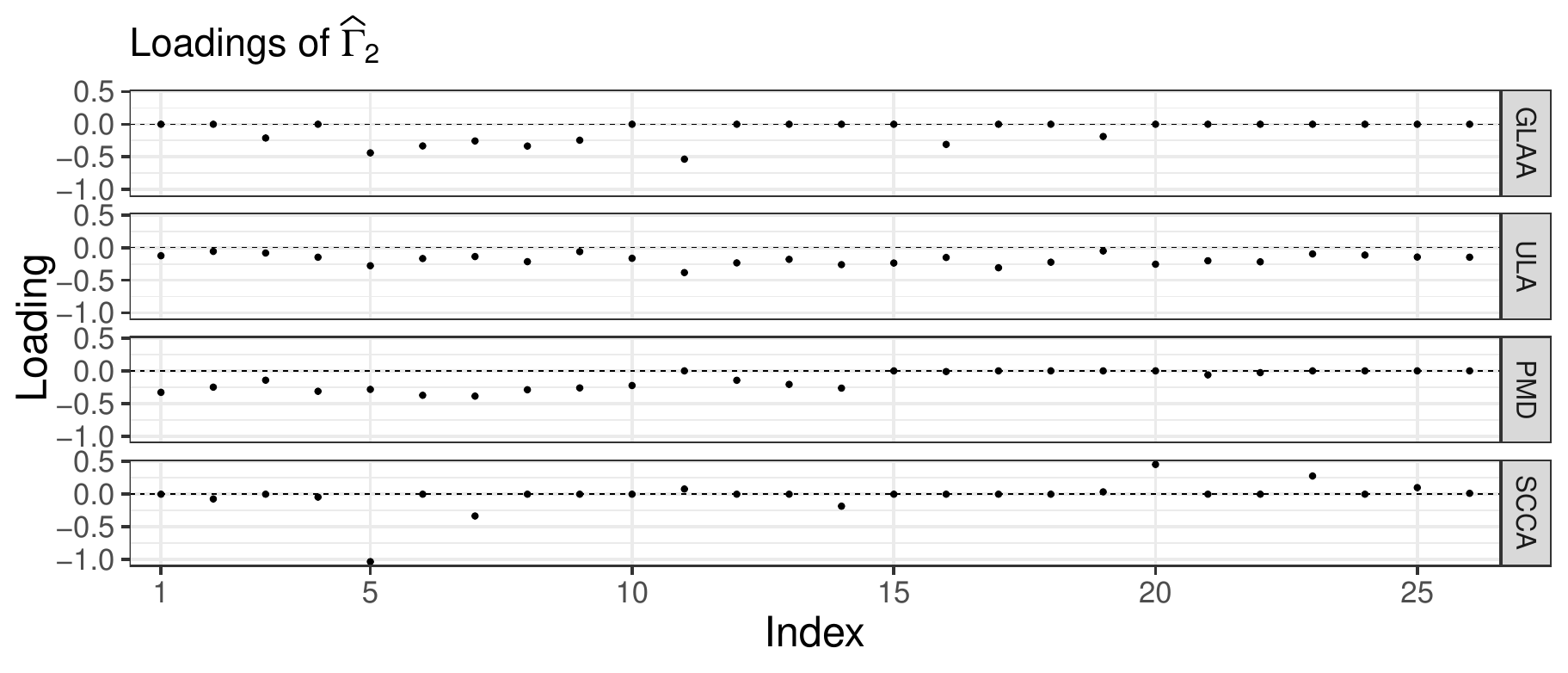}
\end{tabular}
\caption{Estimated loadings in $\hatbolGamma_1$ and $\hatbolGamma_2$. The number of non-zero loading entries estimated by GLAA, ULA, PMD, SCCA are $8, 60, 37, 9$ for $\hatbolGamma_1$, and $9, 26, 16, 11$ for $\hatbolGamma_2$, respectively.}
\label{fig:loading}
\end{figure}

Next, we examine more closely the brain regions identified by GLAA that demonstrate dynamic association patterns. Figure \ref{fig:loading} plots the loadings of the estimated $\hatbolGamma_1$ and $\hatbolGamma_2$, where the indices of non-zero loadings correspond to the selected regions. The number of non-zero loading entries estimated by GLAA, ULA, PMD, SCCA are $8, 60, 37, 9$ for $\hatbolGamma_1$, and $9, 26, 16, 11$ for $\hatbolGamma_2$, respectively. Note that, the ULA method does not deal with variable selection, and for the real data, no information on the true sparsity level is known, so its estimated loadings are non-sparse. Moreover, the PMD method yields a large number of non-zero estimates, making the interpretation difficult. The SCCA method selects about the same number of non-zero regions as GLAA, but the selected regions are less meaningful and are difficult to interpret.

\begin{figure}[t!]
\centering
\resizebox{\textwidth}{!}{
\begin{tabular}{cc}
\includegraphics[width=3.5in]{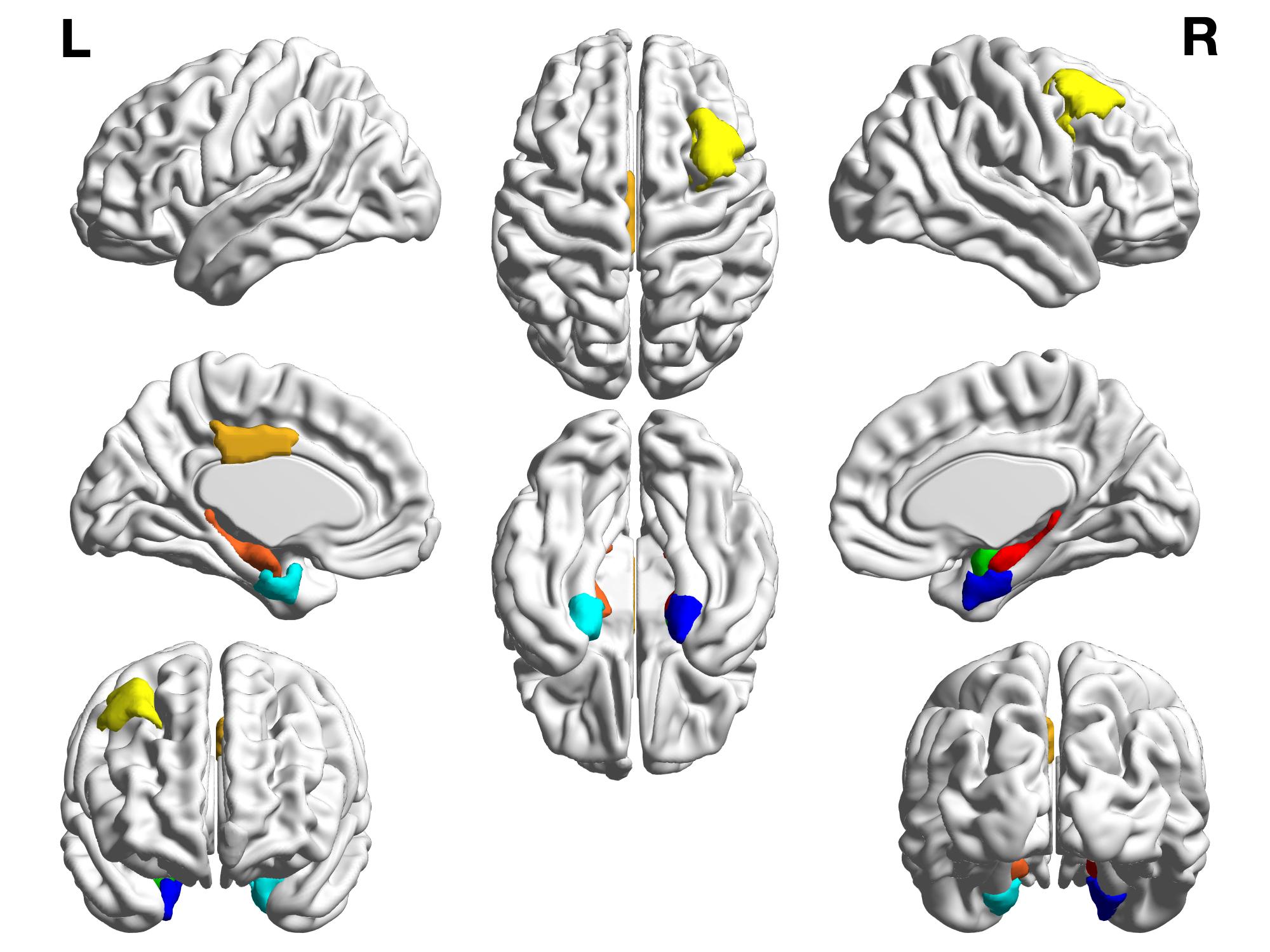} & 
\includegraphics[width=3.5in]{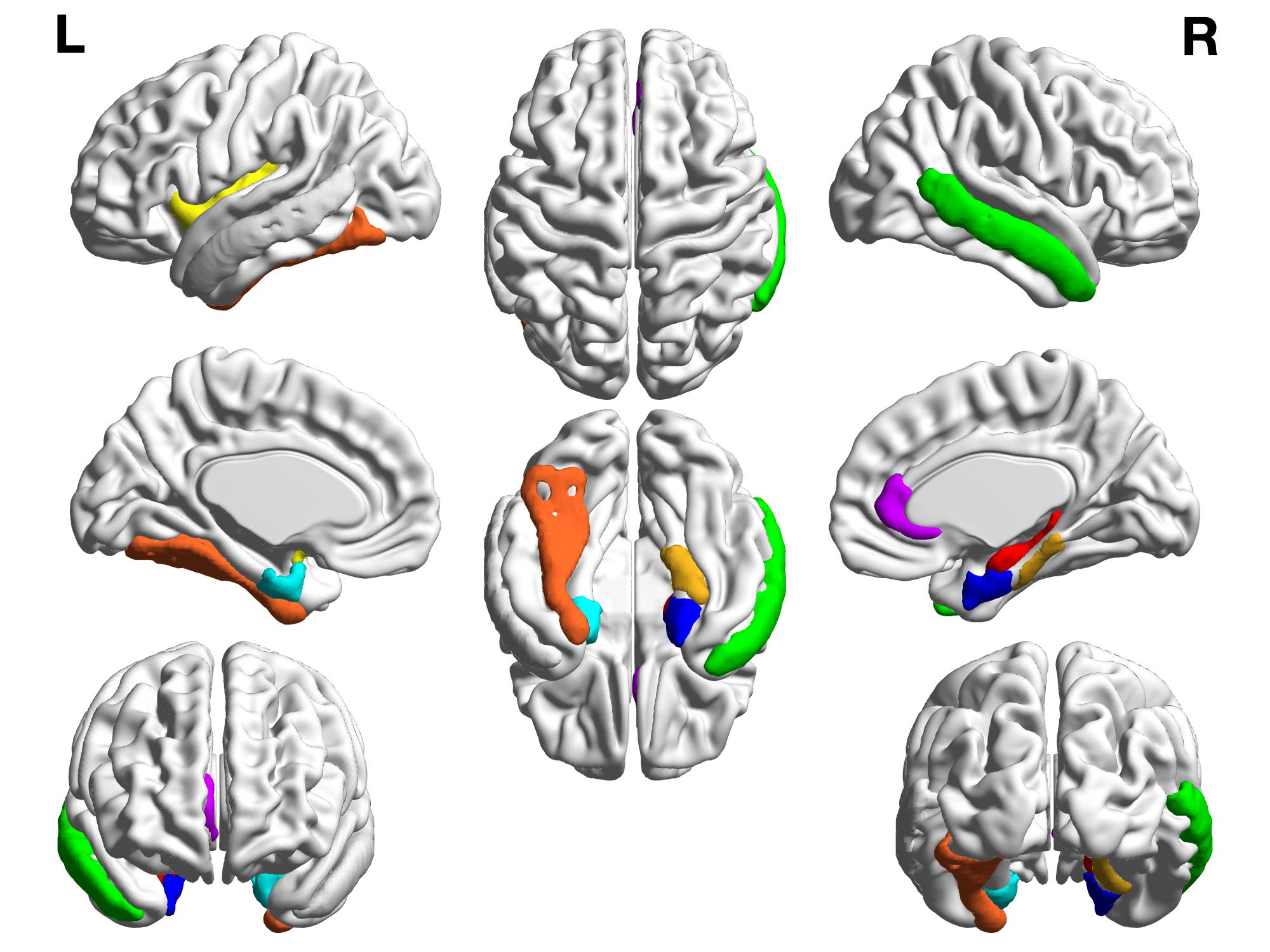} \\
(a) amyloid-beta & (b) tau \\
\end{tabular}
}
\caption{Identified brain regions for amyloid-beta and tau by GLAA.}
\label{fig:bregions}
\end{figure}

Table \ref{tab:bregions} reports the identified brain regions by GLAA for amyloid-beta and tau, respectively, while Figure \ref{fig:bregions} visualizes those regions on a template brain using BrainNet Viewer \citep{Xia2013}. Many of these regions are known to be closely related to AD, and the dynamic associations between amyloid-beta and tau of those regions reveal interesting and new insights. Particularly, for both amyloid-beta and tau, the identified regions include hippocampus and entorhinal cortex. The hippocampus is a major component of the human brain located in the medial temporal lobe, and is functionally involved in response inhibition, episodic memory, and spatial cognition. It is one of the first brain regions to suffer damage from AD, and the hippocampus atrophy is a well-known biomarker for AD \citep{Jack2011}. The entorhinal cortex is a brain area also located in the medial temporal lobes, and functions as a hub in a widespread network for memory, navigation and the perception of time. The entorhinal cortex is the main interface between the hippocampus and neocortex, and together with hippocampus, plays an important role in memories. Atrophy in the entorhinal cortex has been consistently reported in AD \citep{Pini2016}. Moreover, animal models have suggested that neurofibrillary tangles of tau first appear in the entorhinal cortex, then spread to the hippocampus \citep{Cho2016}. For amyloid-beta, other identified regions include amygdala, orbitofrontal cortex, posterior cingulate cortex and areas of middle frontal cortices. The amygdala locates within the temporal lobes of the brain, and performs a primary role in the processing of memory, decision-making and emotional responses. Amygdala atrophy is found prominent in early AD \citep{Poulin2011}. The orbitofrontal cortex locates in the frontal lobes of the brain, and is involved in the cognitive process of decision-making, while the posterior cingulate cortex is part of the cingulate cortex, is one of the most metabolically active brain regions, and is linked to emotion and memory. Atrophy of both regions and the middle frontal cortices have all been found associated with AD \citep{Hoesen2000, Minoshima1997, Pini2016}. For tau, other identified regions include parahippocampal gyrus, middle temporal gyrus, fusiform, insula, and rostral anterior cingulate cortex. The parahippocampal gyrus is a region surrounding the hippocampus, and plays an important role in memory encoding and retrieval. Atrophy in the parahippocampal gyrus has been identified as an early biomarker of AD \citep{Echavarri2011}. Middle temporal gyrus is located on the temporal lobes, and is connected with processes of recognition of known faces and accessing word meaning while reading. The fusiform is located above the parahippocampal gyrus, and is linked with various neural pathways related to recognition. The insula is a portion of the cerebral cortex folded deep within the lateral sulcus, and is  involved in consciousness and diverse functions linked to emotion. The rostral anterior cingulate cortex is frontal part of the cingulate cortex, and is involved in higher-level functions, such as attention allocation, decision-making and emotion. There have been evidences suggesting the associations between these regions and AD \citep{Convit2000, Pini2016}.

\begin{table}[t!]
\centering
\resizebox{\textwidth}{!}{
\begin{tabular}{c|lllll} \toprule
Modality & \multicolumn{5}{c}{Identified regions} \\ \hline
\multirow{2}{*}{amyloid-beta} & Entorhinal R & Entorhinal L & Hippocampus R & Hippocampus L & Amygdala R \\ 
 & Orbitofrontal L & Posterior Cingulate L & Middle Frontal R \\ \hline
\multirow{2}{*}{tau} & Entorhinal R & Entorhinal L & Hippocampus R & Parahippocampal R & Fusiform L \\ 
 & Middle Temporal R & Middle Temporal L & Insula L & Rostral Anterior Cingulate R \\ \bottomrule
\end{tabular}
}
\caption{Identified brain region names for amyloid-beta and tau by GLAA. Regions in the left hemisphere are denoted by ``L", and regions in the right hemisphere are denoted by ``R".} 
\label{tab:bregions}
\end{table}

In summary, GLAA identifies interesting dynamic association patterns among a number of important brain regions between amyloid-beta and tau as age increases. Moreover, GLAA provides a useful dimension reduction tool to help visualize such patterns.

%%%%%%%%%%%%%%%%%%%%%%%%%%%%%%%%%%%%%%%%%%%%%%%%%%%
\section{Discussion}
\label{sec:discussion}

In this article, we propose generalized liquid association analysis, which offers a new angle to study three-way associations among random variables, and is particularly useful for multimodal integrative data analysis. Next, we discuss some potential extensions.

First, we begin with the situation when there is a univariate and categorical $Z$, whereas the analysis so far has primarily concentrated on the case when each variable in $\mbZ$ is continous. In general, it remains an open question on how to define liquid association for a categorical variable, since the function $g(z)$ is no longer differentiable for a categorical $Z$. For a binary $Z\in\{0,1\}$, we propose to replace the derivative of the conditional mean function with the absolute change in the conditional means across the two groups, i.e., $\LA(X,Y | Z)=\left|\E(XY | Z=1)-\E(XY | Z=0)\right|$, where the absolute value is used because the class labels are interchangeable. This naturally fits the original interpretation of LA. Similarly, for a categorical or ordinal $Z\in\{1,\dots,K\}$, we can use the weighted sum of pairwise absolute mean difference between the pairs of groups. Accordingly,  the liquid association of $\mbX$ and $\mbY$ given $Z$ is defined as a $p_{1} \times p_{2}$ matrix. 

Next, for a multivariate mixed type $\mbZ$, we first organize $\mbZ=(\mbZ_{1}, \mbZ_{2})^{T}$ to separate the continuous variables, $\mbZ_{1} = (Z_{1},\dots,Z_{q})^T \in \mbbR^{q}$, from the categorical variables, $\mbZ_2 = (Z_{q+1},\dots, Z_{p_{3}})^T$ $\in \mbbR^{p_3-q}$. Directly imposing a low-dimensional structure on entire $\mbZ$ would lead to difficulty in interpretation. Alternatively, we propose a dimension reduction approach, by recognizing the reduction on $\mbZ$ in model \eqref{XYZsub} is indeed a sufficient dimension reduction model. Specifically, when $\mbZ$ is continuous, by model \eqref{XYZsub}, we have $\mbg(\mbZ) = \mbg(\mbP_{\calS}\mbZ)$, where $\calS = \spn(\bolGamma_3)$. Therefore, $\mbg(\mbZ) \independent \mbZ | \mbP_{\calS}\mbZ$. This leads to a sufficient dimension reduction model of $\mbZ$ for the conditional mean function $g(\mbZ) = \E(\mbX\mbY^T | \mbZ)$, in the sense that all the mean information of the regression of the matrix response $\mbX\mbY^T$ given the predictor vector $\mbZ$ is fully captured by the linear combinations $\bolGamma_3^{T} \mbZ$. As such, model \eqref{XYZsub} can be viewed as a generalization of the notion of sufficient mean reduction \citep{cook2002dimension}. Now for the mixed type $\mbZ=(\mbZ_1^T,\mbZ_2^T)^T$, we adopt the idea of partial dimension reduction \citep{partialSDR}, or groupwise dimension reduction \citep{groupSDR}, and estimate the subspace $\calS \subseteq \mbbR^q$ such that $\mbg(\mbZ) \independent \mbZ \mid (\mbP_{\calS}\mbZ_1,\mbZ_2)$.

From the simulation results in Table \ref{tab:p3}, we observe that, the higher the dimension of $\mbZ$, the more challenging the problem becomes. We believe it is possible to employ an alternative modeling strategy such as \citet{chen2011penalized} when the dimension of $\mbZ$ is ultrahigh. This is also true when the scientific interest is to select important variables in $\mbZ$, while our current interest concentrates on selection of variables in $\mbX$ and $\mbY$, but not in $\mbZ$. We leave the pursuit of this line of research as our future work.

\baselineskip=16.9pt
\bibliographystyle{apalike}
\bibliography{ref_liquid}

\end{document}